\documentclass[pra, preprint,showpacs,amsmath,amsfonts,amssymb,aps]
              {revtex4}
\usepackage{latexsym,makeidx,ifthen,calc,longtable,graphics}
\usepackage{bm}

\newcommand{\rqmversion}{Version 3.3 (ePrint v7)}
\newcommand{\rqmtitle}{Foundations of a Spacetime Path Formalism for 
                       Relativistic Quantum Mechanics}
\newcommand{\rqmauthor}{Ed Seidewitz}
\newcommand{\rqmaddress}{14000 Gulliver's Trail, Bowie MD 20720 USA}

\begin{document}
    
    \title{\rqmtitle}
    \preprint{\rqmversion}
    
    \author{\rqmauthor}
    \email{seidewitz@mailaps.org}
    \affiliation{\rqmaddress}
    
    \date{15 November 2006}
    
    
    \begin{abstract}
        Quantum field theory is the traditional solution to the
        problems inherent in melding quantum mechanics with special
        relativity. However, it has also long been known that an
        alternative first-quantized formulation can be given for
        relativistic quantum mechanics, based on the parametrized
        paths of particles in spacetime. Because time is treated
        similarly to the three space coordinates, rather than as an
        evolution parameter, such a spacetime approach has proved
        particularly useful in the study of quantum gravity and
        cosmology. This paper shows how a spacetime path formalism can
        be considered to arise naturally from the fundamental
        principles of the Born probability rule, superposition, and
        Poincar\'e invariance. The resulting formalism can be seen as
        a foundation for a number of previous parametrized approaches
        in the literature, relating, in particular, ``off-shell''
        theories to traditional on-shell quantum field theory. It
        reproduces the results of perturbative quantum field theory
        for free and interacting particles, but provides intriguing
        possibilities for a natural program for regularization and
        renormalization. Further, an important consequence of the
        formalism is that a clear probabilistic interpretation can be
        maintained throughout, with a natural reduction to
        non-relativistic quantum mechanics.
    \end{abstract}
    
    \maketitle
    
 
    \section{Introduction} \label{sect:intro}

The idea of constructing quantum states as a ``sum over histories'' is
well known in the form of the Feynman path integral formulation.
However, this approach is best known in its application to
non-relativistic quantum mechanics \cite{feynman48, feynman65}, in
which particle paths are parametrized by coordinate time. A natural
relativistic generalization is to consider parametrized paths in
four-di\-men\-sion\-al space\-time rather than time-parametrized
paths in three-dimensional space. Feynman himself developed such an
approach, and this conception seems to have informed much of Feynman's
early view of relativistic quantum mechanics \cite{feynman49,
feynman50, feynman51}.

At an even earlier date, Stueckelberg presented a detailed formulation
of relativistic quantum mechanics in terms of parametrized spacetime
paths \cite{stueckelberg41, stueckelberg42}. A number of other authors
(notably \refcites{fock37, nambu50, schwinger51, morette51, cooke68,
horwitz73, collins78, fanchi78, piron78, fanchi83, fanchi93}) have
also developed related approaches involving an invariant ``fifth
parameter'' governing the evolution of a quantum system, though not
necessarily identifying this explicitly as a path parameter.

A key feature of these approaches is that time is treated comparably
to the three space coordinates, rather than as an evolution parameter.
This is particularly applicable to the study of quantum gravity and
cosmology, in which the fundamental equations (such as the
Wheeler-DeWitt equation) make no explicit distinction for the time
coordinate (see, e.g., \refcites{teitelboim82, hartle83, hartle86,
hartle95, halliwell01a, halliwell01b, halliwell02}).

Also, in the infinite-tension limit, string theory reduces to a
worldline formalism for relativistic quantum theory \cite{bern88,
bern91, bern92, strassler92, schmidt93, schmidt94, schubert01}. One
would therefore expect a path formulation of relativistic quantum
mechanics to provide a natural bridge to the typically first-quantized
formulation of string theory.

Despite the promise of the approach, spacetime path formalisms have
often been presented in the literature as simply alternative
formulations of results obtained from the more traditional quantum
field theory formalism. The motivation of the present paper, however,
is to construct a first-quantized spacetime path formalism that can be
considered foundational in its own right. This means that many typical
tools of field theory, such as Hamiltonian dynamics and the Lagrangian
stationary action principle for fields, cannot be assumed to apply
\emph{a priori}.

Instead, we will begin with the fundamental principles of
special-relativistic quantum theory---the Born probability rule,
superposition, and Poincar\'e invariance---and introduce six
additional, physically motivated postulates related to spacetime
paths. (The perhaps even more fundamental question of why quantum
probabilities are given via superpositions of probability amplitudes
will not be addressed here.) Results deduced from these postulates
then provide the basis for further physical interpretation.

Since this formalism is first quantized, particular care is given to
properly handling particles and antiparticles and to developing a
consistent probabilistic interpretation. The result is an approach
that fully deals with the usual issues of negative energies and
negative probabilities, but without necessitating the introduction of
fields as fundamental entities. Rather, fields can be considered to be
simply a convenient formalism for handling multiparticle states. The
present work only discuses massive scalar particles, but the approach
can be extended to handle non-scalar particles (e.g., 
\refcite{horwitz75, horwitz82a}).

\Sect{sect:free} first introduces the formalism for free scalar
particles, culminating in free multiparticle fields.
\Sect{sect:interacting} then extends the formalism to consider
interacting states and scattering. In order to reduce clutter in the
text, certain propositions resulting from purely mathematical, but
somewhat involved, derivation are cited without proof in the main body
of the text, with proofs given in appendices.

Natural units with $\hbar = 1 = c$ are used throughout the following
and the metric has a signature of $(-+++)$.

\section{Free Particles} \label{sect:free}
    
For any path based approach, it is obviously critical to be clear on
what is meant by the term \emph{path}. In the present case, a
\emph{path} for a particle is an arbitrary curve through spacetime,
that is, a continuous (though not necessarily differentiable),
one-dimensional subspace of spacetime. Note that there is no \emph{a
priori} requirement that such a curve is timelike or lightlike.
Indeed, the path may cross arbitrarily forwards and backwards in time.
Since such a path is continuous, there is a one-to-one mapping between
it and some interval of the real numbers. That is, a path may be given
by functions $\qmul$, for $\mu = 0, 1, 2, 3$, of a \emph{path
parameter} $\lambda$.

In this formulation, the path parameter $\lambda$ serves a purpose
similar to that of time in the traditional non-relativistic path
integral approach \cite{feynman65}. For the restricted case of an
everywhere-timelike path, this parameter is analogous to proper time.
For the general case of an unrestricted path, there has been some
debate as to the physical nature of the path parameter (see, for
example, \refcites{fanchi86, horwitz96}). In order not to presuppose
any specific interpretation, we will consider, for each path, all
possible parameterizations of the path.

To do this, choose a fiducial parametrization $s$, say over the
interval $[0,1]$, and define any other parametrization as a
monotonically increasing function $\lambda(s)$. Geometrically, the so
called \emph{lapse multiplier}
\begin{equation*}
    w(s) \equiv \frac{\dif \lambda}{\dif s} > 0
\end{equation*}
then gives an effective length metric $\dl = w(s)\dif s$ for the path,
and the corresponding parametrization $\lambda$ is an \emph{intrinsic
length} measure along the path.

Given this basic conception of a particle path, this section will
review the fundamental postulates required for a path integral
approach, derive the scalar free particle propagator and carefully
consider the corresponding probability interpretation.

\subsection{The Free Particle Propagator} \label{sect:free:propagator}

The fundamental postulate of any spacetime path integral approach is
that a particle's transition amplitude between two points in spacetime
is a superposition of the transition amplitudes for all possible paths
between those points. Let the functional $\propsym[q]$ give the
transition amplitude for a path $\qmu(\lambda)$. Then the total
transition amplitude $\propsym(x,\xz)$ must be given by a path
integral over $\propsym[q]$, for all paths $q$ from $\xz$ to $x$.

\begin{postulate}
    For a free scalar particle, the transition amplitude
    $\propsym(x,\xz)$ is given by the superposition of path transition
    amplitudes $\propsym[q]$, for all possible 4-dimensional path
    functions $\qmul$ beginning at $\xz$ and ending at $x$,
    parametrized by all possible monotonically increasing functions
    $\lambda(s)$. That is,
    \begin{equation} \label{eqn:G2}
        \propsym(x,\xz) =
            \intD \lambda\,  
                \theta\left[ \frac{\dif \lambda}{\dif s} \right]
                F[\lambda] \kernel \,,
    \end{equation}
    where the real-valued functional $F[\lambda]$ allows for the
    possibility of different weights for different parametrizations,
    and
    \begin{equation} \label{eqn:G3}
        \kernel \equiv
            \eta[\lambda]
            \intDfour q\, 
                \delta^{4}(q(\lambda(1)) - x)
                \delta^{4}(q(\lambda(0)) - \xz)
                \propsym[q] \,,
    \end{equation}
    where $\eta[\lambda]$ is a parametrization-dependent normalization
    factor as required to keep the path integral finite.
\end{postulate}

Note that, in \eqns{eqn:G2} and \eqref{eqn:G3}, the notation
$\Dif\lambda$ indicates a path integral over the parameterization
function $\lambda(s)$ while $\Dfour q$ indicates a path integral over
the four path functions $\qmul$.

In the traditional Feynman sum-over-paths approach, the form of
$\propsym[q]$ is simply assumed to be an exponential of the classical
action \cite{feynman65}. This is justified because the resulting
transition amplitudes agree with the results of the usual formulation
of quantum mechanics. However, if the path-based formulation is to be
considered foundational, one would prefer a more fundamental
justification.

As a transition amplitude, $\propsym[q]$ strictly only applies to a
particle on a specific path $q$ from the starting position
$q(\lambdaz)$ to the ending position $q(\lambda_{1})$ (where the
parameter range of $q$ is $[\lambdaz,\lambda_{1}]$). However, by
translational invariance in Minkowski spacetime, the particle
propagation embodied in $\propsym[q]$ cannot depend on the absolute
positions $q(\lambda)$, but only on the relative positions
\begin{equation*}
    \Delta q (\lambda) \equiv q(\lambda) - q(\lambdaz) \,.
\end{equation*}
That is, we can take $\propsym[q] = \propsym[\Delta q]$.

Now, consider a \emph{family} of parallel paths $q_{\xz}$, indexed by
the starting position $\xz$, such that
\begin{equation*}
    q_{\xz}(\lambda) = \xz + \Delta q(\lambda) \,,
\end{equation*}
for a fixed relative position function $\Delta q$. Since all members
of such a family have the same relative position function $\Delta q$,
the amplitude $\propsym[q_{\xz}] = \propsym[\Delta q]$ must be the
same for all members of the family.

Suppose that a probability amplitude $\psixz$ is given for a particle
to be at an initial position $\xz$ and that the transition amplitude
is known to be $\propsym[\Delta q]$ for a specific relative position
function $\Delta q$. Then, the probability amplitude for the particle
to traverse a specific path $q_{\xz}$ from the family for relative
position $\Delta q$ is just $\propsym[q_{\xz}]\psixz = \propsym[\Delta
q]\psixz$.

However, the very meaning of being on the specific path $q_{\xz}$ is
that the particle must propagate from the starting position at $\xz$
to the ending position at $q_{\xz}(\lambda_{1})$. Therefore, the
probability for reaching the end position $q_{\xz}(\lambda_{1})$ must
be the same as the probability for having started out at the position
$\xz$. That is,
\begin{equation*}
    \sqr{\propsym[\Delta q]\psixz} = \sqr{\psixz} \,.
\end{equation*}
But, since $\propsym[\Delta q]$ is independent of $\xz$, we must have 
$\sqr{\propsym[q]} = 1$ in general.

Of course, this argument is really just a suggestive motivation rather
than a proof, so we take the conclusion as a postulate, rather than a
proposition.

\begin{postulate}
    For any path $\qmul$, the transition amplitude $\propsym[q]$
    preserves the position probability density for the particle along
    the path. That is, it satisfies
    \begin{equation} \label{eqn:A0a}
        \sqr{\propsym[q]} = 1 \,.
    \end{equation}
    
\end{postulate}
If the configuration space for a path is expanded to be a
representation of the full Poincar\'e group---that is, to include a
matrix representation of the (homogeneous) Lorentz group as well as
the four spacetime coordinates---then members of a family of
``parallel'' paths are related by Poincar\'e transformations, not just
translations. This can be used as the basis for extending the
spacetime path formalism to cover non-scalar particles. If, further,
the assumption of flat spacetime is dropped, then it is not generally
possible to construct a family of parallel paths covering all
spacetime. However, one can still consider infinitesimal variations
along a path corresponding to arbitrary coordinate transformations.
Such further generalizations of the spacetime path approach will be
explored in future papers.

The requirements of \eqn{eqn:A0a} and translation invariance mean
that $\propsym[q]$ must have the form
\begin{equation} \label{eqn:A2}
    \propsym[q] = \me^{\mi S[\Delta q]} \,,
\end{equation}
for some phase functional $S$. Substituting \eqn{eqn:A2} into
\eqn{eqn:G3} gives
\begin{equation} \label{eqn:A2a}
    \kernel =
        \eta[\lambda] 
         \intDfour q\, 
             \delta^{4}(q(\lambda(1)) - x)
             \delta^{4}(q(\lambda(0)) - \xz)
             \me^{\mi S[\Delta q]} \,.
\end{equation}

So far, we have made no assumption that the particle path functions
$\qmul$ are differentiable. Indeed, paths under a path integral will
generally \emph{not} be differentiable. Nevertheless, it is common
practice to use (with some care) path derivatives in the integrand of
a path integral. This is because a path integral is defined as the
limit of discretized approximations in which path derivatives are
approximated as the mean value $\Delta q / \Delta \lambda$, for finite
differences $\Delta q$ and $\Delta \lambda$. The limit $\Delta\lambda
\to 0$ is then taken over the path integral as a whole, not each
derivative individually. Thus, even though $\lim_{\Delta\lambda \to
0}\Delta q/\Delta \lambda$ may not be defined, the path integral has a
well-defined value so long as the overall path integral limit is
defined. (For a discussion of some of the issues involved here, see,
for example, Section 7.3 of \refcite{feynman65}. See also the explicit
example of the derivation in \app{app:path}.)

We are therefore justified in replacing the difference functions
$\Delta \qmul$ used in the phase functional under the path integral in
\eqn{eqn:A2a} with the path derivatives $\qdotmul \equiv \dif\qmu /
\dif\lambda$, such that
\begin{equation*}
    \Delta \qmul = \int_{\lambdaz}^{\lambda} \dif\lambda' \,
                   \qdotmu(\lambda') \,,
\end{equation*}
letting the $\qmul$ be considered as differentiable. This gives
\begin{equation} \label{eqn:B}
    \kernel =
        \eta[\lambda]
        \intDfour q\, 
            \delta^{4}(q(\lambda(1)) - x)
            \delta^{4}(q(\lambda(0)) - \xz)
            \me^{\mi S[\qdot]} \,.
\end{equation}
which reflects the typical form of a Feynman sum over paths
\cite{feynman65}, where each path is weighted by a phase determined by
the action $S$. Unlike the usual non-relativistic formulation,
however, the path parameter here is $\lambda$, rather than time
\cite{feynman51,teitelboim82}.

Now, by dividing a path $q$ into two paths at some arbitrary parameter
value $\lambda$ and propagating over each segment, we can see that
\begin{equation} \label{eqn:B1}
    S[\qdot;\lambda_{1},\lambdaz] = S[\qdot;\lambda_{1},\lambda] +
                                Ê   S[\qdot;\lambda,\lambdaz] \,,
\end{equation}
where $S[\qdot;\lambda',\lambda]$ denotes the value of $S[\qdot]$ for
the parameter range of $\qdot$ restricted to $[\lambda,\lambda']$.
Using this property to build the total value of $S[\qdot]$ from
infinitesimal increments leads to the following result (proved in
\app{app:phase}).

\begin{theorem}[Form of the Phase Functional]
    The phase functional $S$ must have the form
    \begin{equation} \label{eqn:D1}
        S[\qdot] = \int^{\lambda_{1}}_{\lambdaz} \dl' \, 
                       L[\qdot;\lambda'] \,,
    \end{equation}
    where the parametrization domain for $\qdot$ is
    $[\lambdaz,\lambda_{1}]$ and $L[\qdot;\lambda]$ depends only on
    $\qdot$ and its higher derivatives evaluated at $\lambda$.
\end{theorem}


The question remains as to what form the function $L$ should take.
Traditionally, it is taken to be just the classical Langrangian, but,
from a foundational viewpoint, one would like a better justification.

Of course, the simplest form for $L$ would be a constant, independent
of $\qdot$. However, this would result in a superficially divergent
path integral in \eqn{eqn:B} which, when normalized, would leave to
just a trivial phase. This would not give any appropriate particle
dynamics. The next simplest form for $L$ would be for it to depend
only on $\qdot$ and no higher derivatives. Further, since $L$ is a
scalar quantity, it must then depend only on the Lorentz-invariant
scalar function
\begin{equation*}
     \qdotsql \equiv \qdotmul \qdot_{\mu}(\lambda) \,.
\end{equation*}

Taking $L$ to further have the tractable form of a linear function of 
$\qdotsq$ gives
\begin{equation*}
    L[\qdot; \lambda] = L(\qdotsql) = a\qdotsql + b \,,
\end{equation*}
for some $a$ and $b$. Now, the factor $a$ can be fixed arbitrarily,
since any variation is effectively equivalent to a reparametrization
of the path parameter $\lambda$. For a free particle, it is convenient
to take $a = 1/4$. If we further assume that $b$ is always negative, 
we can set $b = -m^{2}$ and identifying $m$ with the mass of the 
particle does, indeed, give a classical relativistic Lagrangian 
function.

As we will see in the following, evaluating the path integral in
\eqn{eqn:B} with this Lagrangian function leads to the usual
free-particle Feynman propagator for the particle. If, on the other
hand, we take $b$ to be positive, then the result is a similar
propagator, but with an effective imaginary particle mass. Such
particles are tachyons, which we will not consider further in this
paper.

\begin{postulate}
    For a free scalar particle of mass $m$, the Lagrangian function is
    given by
    \begin{equation} \label{eqn:E}
        L(\qdotsq) = \frac{1}{4}\qdotsq - m^{2} \,.
    \end{equation}
\end{postulate}

Substituting \eqn{eqn:D1} into \eqn{eqn:B} gives
\begin{equation} \label{eqn:F1}
    \kernel =
        \eta[\lambda]
        \intDfour q\, 
            \delta^{4}(q(\lambda(1)) - x)
            \delta^{4}(q(\lambda(0)) - \xz)
                \exp \left(
                    \mi\int^{\lambda(1)}_{\lambda(0)}\dl' \,
                        L(\qdotsq(\lambda'))
                \right) \,.
\end{equation}
With the Lagrangian given by \eqn{eqn:E}, it is well known that this
path integral may be evaluated (see, for example,
\refcite{teitelboim82}). However, in the present context, some care
must be taken to mathematically evaluate the integral without making
any further assumptions based on field equations or underlying
traditional quantum mechanics. In any case, the result (proved in
\app{app:path}) is as follows.

\begin{theorem}[Evaluation of the Path Integral]
    The path integral in \eqn{eqn:F1}, with the Lagrangian given by
    \eqn{eqn:E}, may be evaluated to get
    \begin{equation} \label{eqn:H}
        \kernel = \kersym(x-\xz;\lambda(1)-\lambda(0))
            \equiv (2\pi)^{-4} \intfour p \, \me^{\mi p\cdot(x - \xz)}
                   \me^{-\mi [\lambda(1)-\lambda(0)](p^{2} + m^{2})} \,.
    \end{equation}
    
\end{theorem}
Note that the only dependency left of
$\kersym(x-\xz;\lambda(1)-\lambda(0))$ on the parametrization
$\lambda(s)$ is on the total \emph{intrinsic path length}
\begin{equation*}
    T = \lambda(1) - \lambda(0) = \int_{0}^{1} \dif s\, w(s) > 0 \,.
\end{equation*}
If we were to take $F[\lambda] = 1$ in \eqn{eqn:G2} for all
$\lambda(s)$, there would then be a parametrization gauge symmetry:
all parametrizations that give the same intrinsic path lengths would
be equivalent. Therefore, equivalent reparametrizations would be
overcounted in the $\lambda$ path integral of \eqn{eqn:G2}, so the
integral would diverge. 

If, on the other hand, the path integral over $\lambda$ had not been
included at all in \eqn{eqn:G2}, the result would have been to
overspecify a specific path parametrization. The possible particle
paths would then have been \emph{undercounted}, missing the need to
include paths of different intrinsic lengths. It is thus necessary to
reduce the $\lambda$ path integration in \eqn{eqn:G2} to eliminate the
overcounting due to the path gauge symmetry, without overspecifying
the path parametrization.

In the usual fashion for a gauge symmetry, we retain the integration,
but fix a specific guage. This can be easily done by including a gauge
fixing delta functional in $F[\lambda]$. The gauge typically chosen is
to require that $w(s) = \dif\lambda/\dif s$ be constant
\cite{teitelboim82}, which corresponds to setting
\begin{equation*}
    F[\lambda] = f(\lambda(1)-\lambda(0))
                 \delta \left[
                     \frac{\dif\lambda}{\dif s} - 
                     [\lambda(1)-\lambda(0)]
                 \right] \,,
\end{equation*}
for some real function $f(T)$. Using this in \eqn{eqn:G2} gives
\begin{equation*}
    \propsym(x,\xz) = \int_{0}^{\infty} \dif T\, f(T) \kersym(x-\xz; T) \,.
\end{equation*}

In the following, we will generally assume equal weighting of all
parametrizations, that is $f(T) = 1$. However, in
\sect{sect:interacting:renormalization}, we will see that an alternate
choice provides a fruitful path for regularizing the infinite
integrals that appear in the formalism for interacting particles.
Nevertheless, assuming, for now, that $f(T) = 1$, gives
\begin{equation} \label{eqn:I2a}
    \propsym(x, \xz)
        = \int_{0}^{\infty} \dif T\, \kersym(x-\xz; T)
        = (2\pi)^{-4} \intfour p\, \me^{\mi p\cdot(x-\xz)}
              \int_{0}^{\infty} \dif T\, \me^{-\mi T(p^{2}+m^{2})} \,.
\end{equation}
This can be evaluated by introducing a convergence factor
$\exp(-T\varepsilon)$, for infinitesimal $\varepsilon$, resulting in
just the Feynman propagator
\begin{equation*} 
    \propsym(x,\xz) = \prop 
        \equiv -\mi(2\pi)^{-4}\intfour p\, 
               \frac{\me^{\mi p\cdot(x-\xz)}}
                   {p^{2}+m^{2}-\mi\varepsilon} \,.
\end{equation*}

The integration of $T$ from $0$ to $\infty$ in \eqn{eqn:I2a} is
similar to the integration carried out by Nambu \cite{nambu50}, based
on previous work of Fock \cite{fock37}, in order to obtain the Feynman
propagator. Note, though, that this integration arises naturally here
as the guage-fixed reduction of the path parametrization integral in
\eqn{eqn:G2}.

The relationship between the propagator $\prop$ and $\kersym(x-\xz;T)$
can be viewed in another way, which will also prove useful in
\sect{sect:interacting:renormalization}. For $T > 0$,
\begin{equation} \label{eqn:I3}
    \begin{split}
        \kersym(x-\xz;T)
            &= \me^{-\mi T m^{2}} 
               \intfour p\, \me^{\mi p\cdot(x-\xz)}
               \int_{0}^{\infty} \dif T'\, \me^{-\mi T' p^{2}}
               \delta(T'-T) \\
            &= (2\pi)^{-1}\me^{-\mi T m^{2}}
               \intfour p\, \me^{\mi p\cdot(x-\xz)}
               \int_{0}^{\infty} \dif T'\, \me^{-\mi T' p^{2}}
               \int \dif m'^{2}\,
               \me^{-\mi(T'- T)m'^{2}} \\
            &= (2\pi)^{-1}\me^{-\mi T m^{2}}
               \int \dif m'^{2}\, \me^{\mi T m'^{2}}
               \propsym(x-\xz;m'^{2}) \,,
    \end{split}
\end{equation}
where
\begin{equation}
    \propsym(x-\xz;m'^{2}) 
        \equiv \int_{0}^{\infty} \dif T'\,
                   \intfour p\, \me^{\mi p\cdot(x-\xz)}
                       \me^{-\mi T'(p^{2}+m'^{2})}
        =      -\mi(2\pi)^{-4}\intfour p\, 
                   \frac{\me^{\mi p\cdot(x-\xz)}}
                        {p^{2}+m'^{2}-\mi\varepsilon} \,.
\end{equation}
This form for $\kersym(x-\xz;T)$ is essentially that of the
parametrized Green's function derived by Horwitz et al.\ for
parametrized quantum field theory \cite{land91,frastai95} as a
superposition of propagators for different mass states (see also
\refcite{enatsu63,enatsu86}). \Eqn{eqn:I3} differs from those
references in the factor $\exp(-\mi T m^{2})$. As a result of this
factor, integrating \eqn{eqn:I3} over $T$ as in \eqn{eqn:I2a}
effectively acts as a Fourier transform, resulting in a propagator
with mass sharply defined at $m$.

\subsection{Free Particle Position States} \label{sect:free:states}

The path integral form for $\kerneld$ given in \eqn{eqn:G3} is
essentially the same as that of the path integral for the
non-relativistic \emph{kernel} \cite{feynman65}, except that $\lambda$
is used as the evolution parameter instead of $t$. Therefore,
$\kerneld$ has similar properties as a propagation kernel in
$\lambda$:
\begin{equation*}
    \intfour x_{1}\, \kersym(x-x_{1}; \lambda-\lambda_{1})
                     \kersym(x_{1}-\xz; \lambda_{1}-\lambda_{0})
        = \kerneld
\end{equation*}
and
\begin{equation*}
    \kerneld^{*} = \kersym(\xz - x; \lambdaz - \lambda) \,.
\end{equation*}

Given these properties, define a \emph{family} of probability
amplitude functions $\psixl$, for which
\begin{equation} \label{eqn:F2}
    \psixl = \intfour \xz\, \kerneld \psixlz \,,
\end{equation}
for any $\lambda$ and $\lambdaz$, normalized such that
\begin{equation} \label{eqn:A0}
    \intfour x\, \sqr{\psixl} = 1 \,,
\end{equation}
for each $\lambda$. Formally, these functions are probability
amplitudes for the position $x$, with $\lambda$ serving as an index
identifying individual functions in the family. However, they can be
interpreted as just the parametrized probability amplitude
functions defined by Stueckelberg \cite{stueckelberg41}. In this
sense, the $\psixl$ represent the probability amplitude for a particle
to reach position $x$ at the point along its path with parameter value
$\lambda$.

Note that
\begin{equation*}
    \mi \pderiv{}{\lambda} \kerneld
        = (2\pi)^{-4} \intfour p \, \me^{\mi p\cdot(x - \xz)}
          (p^{2} + m^{2})
          \me^{-\mi(\lambda - \lambdaz)(p^{2} + m^{2})} \,.
\end{equation*}
This means that $\psixl$, as given by \eqn{eqn:F2}, satisfies
\begin{equation} \label{eqn:H0}
    -\mi \pderiv{}{\lambda} \psixl 
        = \left( 
            \frac{\partial^{2}}{\partial x^{2}} - m^{2}
          \right) \psixl \,.
\end{equation}
\Eqn{eqn:H0} is a generalized Schr\"odinger equation, such as proposed
by Stueckelberg \cite{stueckelberg42}. However, Stueckelberg and
subsequent authors \cite{feynman50, horwitz73, fanchi78} used a
Hamiltonian of the form $(2m)^{-1}\partial^{2}/\partial x^{2}$, by
analogy with non-relativistic mechanics, rather than the form of
\eqn{eqn:H0} (though \refcite{horwitz81} uses a Hamiltonian form
similar to \eqn{eqn:H0}). This difference is the origin of the extra
factor $\exp(-\mi T m^{2})$ in \eqn{eqn:I3} relative to
\refcites{land91,frastai95}.

The properties of the kernel $\kerneld$ also allow for the definition
of a consistent family of \emph{position state} bases $\ketxl$, such
that
\begin{equation} \label{eqn:A}
    \psixl = \innerxlpsi \,,
\end{equation}
given a single Hilbert space state vector $\ketpsi$. These position
states are normalized such that
\begin{equation*}
    \inner{x'; \lambda}{x; \lambda} = \delta^{4}(x' - x) \,.
\end{equation*}
for each value of $\lambda$. Further, it follows from \eqns{eqn:F2} and
\eqref{eqn:A} that
\begin{equation} \label{eqn:F0}
    \kerneld = \innerxlxlz \,.
\end{equation}
Thus, $\kerneld$ effectively defines a unitary transformation between 
the various Hilbert space bases $\ketxl$, indexed by the parameter
$\lambda$.

Finally, the overall state for propagation from $\xz$ to $x$ is given
by the superposition of the states for paths of all intrinsic lengths.
If we fix $\qmulz = \xmu_{0}$, then $\ketxl$ already includes all
paths of length $\lambda - \lambdaz$. Therefore, the overall state
$\ketx$ for the particle to arrive at $x$ should be given by the
superposition of the states $\ketxl$ for all $\lambda > \lambdaz$:
\begin{equation} \label{eqn:I2}
    \ketx \equiv \int_{\lambdaz}^{\infty} \dl\, \ketxl \,.
\end{equation}
Then, using \eqn{eqn:F0},
\begin{equation} \label{eqn:I2b}
    \innerxxlz
           = \int_{\lambdaz}^{\infty} \dl\, \kerneld 
           = \int_{0}^{\infty} \dl\, \kersym(x-\xz; \lambda)
           = \prop \,.
\end{equation}

Now, the $\ketx$ are not actually proper Hilbert space states, since
$\inner{x}{\xz}$ is infinite (as can be see by integrating
\eqn{eqn:I2b} over $\lambdaz$). Nevertheless, via \eqn{eqn:I2a}, the
corresponding bras $\brax$ can be considered to be well-defined
functions on proper, normalizable states $\ket{\psi}$ such that
\begin{equation*}
    \inner{x}{\psi} = \intfour \xz\, \prop\psixlz
\end{equation*}
is the transition amplitude for a particle with known probability
amplitude $\psixlz$ at $\lambdaz$ to eventually reach position $x$ at 
some $\lambda > \lambdaz$. We will thus continue to use $\ketx$ as a
formal quantity, with the understanding that it is really just a
shorthand for constructing propagators and transition amplitudes.

\subsection{On-Shell Particle and Antiparticle States}
\label{sect:free:onshell}

The states constructed so far have naturally been off-shell states.
That is, they represent what are normally considered to be ``virtual''
particles. However, rather than simply imposing the on-shell mass
condition to obtain ``physical'' states, on-shell states will be
constructed in this subsection as the infinite time limit of off-shell
states. That is, particles with paths that, in the limit, are
unbounded in time will turn out to be naturally on-shell.

In order to take a time limit, it is necessary to make some
distinction between past and future that can be used as the basis for
taking the limit. For this purpose, divide the set of all possible
paths $\qmu$ that end at some specific $\qmul = \xmu$ into two
subsets: those that begin (at $\qmulz = \xmu_{0}$) in the \emph{past}
of $x$ and those that begin in the \emph{future} of $x$. 

Outside of the light cone of $x$, the division into future and past
is, of course, not Lorentz covariant and depends on the choice of a
specific coordinate system. However, when we take the time limit, the
light cone expands to cover all space, and, in this limit, the
division into particle and antiparticle becomes fully coordinate
system independent. The possibility of the particle/antiparticle 
distinction being coordinate system dependent in anything other than 
the infinite time limit is a subject for future exploration.

Now, particles are normally considered to propagate \emph{from} the
past \emph{to} the future. On the other hand, \emph{antiparticles} may
be considered to propagate from the \emph{future} into the \emph{past}
\cite{stueckelberg41, stueckelberg42, feynman49}.

\begin{postulate}
    Normal particle states $\ketax$ are such that
    \begin{equation*}
        \innerxaxlz = \thetaax \prop \,.
    \end{equation*}
    Antiparticle states $\ketrx$ are such that
    \begin{equation*}
        \innerxrxlz = \thetarx \prop \,.
    \end{equation*}
\end{postulate}

Using the usual decomposition of the Feynman propagator (see, for
example, Section 6.2 of \refcite{weinberg95})
\begin{equation} \label{eqn:K1}
    \prop = \thetaax \propa + \thetarx \propr\,,
\end{equation}
where
\begin{equation} \label{eqn:L}
    \propar \equiv (2\pi)^{-3} \intthree p\,
                   \frac{\me^{\mi[\mp\Ep(x^{0}-\xz^{0}) +
                              \threevec{p}\cdot
                                  (\threevec{x}-\threevec{x}_{0})]}}
                        {2\Ep} \,,
\end{equation}
with $\Ep \equiv \sqrt{\threevec{p}^{2} + m^{2}}$, it is clear that
\begin{equation} \label{eqn:L1}
    \innerxarxlz = \thetaarx \prop = \thetaarx \propar \,.
\end{equation}

We would now like to take the time limits for future and past directed
particle and antiparticle states. In doing this, one cannot expect to
hold the 3-position of the path end point constant. However, for a
free particle, it is reasonable to take the particle \emph{3-momentum}
as being fixed. Therefore, consider the state of a particle or
antiparticle with a 3-momentum $\threep$ at a certain time $t$. (The
importance of the specific factor $\exp(\mp\mi\Ep t)$ in the
definition below will become clear in a moment.)

\begin{postulate}
    The state of a particle ($+$) or antiparticle ($-$) with
    3-momentum $\threep$ is given by
    \begin{equation} \label{eqn:N1}
        \begin{split}
            \ketartp &\equiv (2\pi)^{-3/2} \intthree x \, 
                            \me^{\mi(\mp\Ep t + \threep\cdot\threex)} 
                            \ketartx \\
                     &=     (2\pi)^{-1/2} \me^{\mp\mi\Ep t} 
                            \int \dif p^{0}\, \me^{\mi p^{0}t} \ketarp
                            \,,
        \end{split}
    \end{equation}
    where
    \begin{equation} \label{eqn:L3}
        \ketarp \equiv (2\pi)^{-2} \intfour x\, \me^{\mi p \cdot x} \ketarx
    \end{equation}
    is the corresponding 4-momentum state.
\end{postulate}
 
Let
\begin{equation} \label{eqn:N2}
    \begin{split}
        \ketarlz{t_{0}, \threep}
            &\equiv (2\pi)^{-3/2} \intthree x\, 
                    \me^{\mi(\mp\Ep t_{0} + \threep\cdot\threex)} 
                    \ketlz{t_{0}, \threex} \\
            &=      (2\pi)^{-1/2} \me^{\mp\mi\Ep t_{0}}
                    \int \dif p^{0}\, \me^{\mi p^{0}t_{0}} \ketplz \,,
    \end{split}
\end{equation}
where
\begin{equation} \label{eqn:J1}
    \ketplz \equiv (2\pi)^{-2} \intfour x\, \me^{\mi p \cdot x} 
                                            \ketlz{x} \,.
\end{equation}
Substituting from \eqns{eqn:L3}, \eqref{eqn:J1} and \eqref{eqn:L1},
\begin{equation} \label{eqn:L4}
    \begin{split} 
        \innerparplz &= (2\pi)^{-4} \intfour x\, \dfour \xz\, 
                        \me^{-\mi p \cdot x} \me^{\mi \pz \cdot \xz}
                        \thetaarx \propar \\
                     &= \delta^{4}(p - \pz) \proparp \,,
    \end{split}
\end{equation}
where
\begin{equation} \label{eqn:M}
    \proparp \equiv \intfour x\, \me^{-\mi p \cdot x} \theta(\pm x^{0}) 
                    \proparx \,.
\end{equation}
Substituting \eqn{eqn:L} into \eqn{eqn:M} gives
\begin{equation} \label{eqn:N}
    \begin{split}
        \proparp &= \intfour x\, \me^{-\mi p \cdot x} \theta(\pm x^{0})
                    (2\pi)^{-3} \intthree p'\, (2\Epp)^{-1}
                    \me^{\mi(\mp\Epp x^{0}+\threepp\cdot\threex)} \\
                 &= \int \dt\, \me^{\mi p^{0}t} \theta(\pm t) 
                    \intthree p'\, (2\Epp)^{-1} \me^{\mp \mi\Epp t} (2\pi)^{-3}
                    \intthree x\, \me^{\mi(\threep\,'-\threep)\cdot\threex} \\
                 &= \int \dt\, \me^{\mi p^{0}t} \theta(\pm t) 
                    \intthree p'\, (2\Epp)^{-1} \me^{\mp \mi\Epp t} 
                    \delta^{3}(\threep\,' - \threep) \\
                 &= (2\Ep)^{-1} \int \dt\, \theta(\pm t)
                    \me^{\mi(p^{0} \mp \Ep)t} \,.
    \end{split}
\end{equation}

Using \eqn{eqn:L4} (and the completeness of the $\ketplz$ states) in
\eqn{eqn:N1}, and substituting from \eqn{eqn:N} for $\proparp$, then
gives
\begin{equation*}
    \begin{split}
        \ketartp
            &= (2\pi)^{-1/2} \me^{\mp\mi\Ep t} 
               \int \dif p^{0}\, \me^{\mi p^{0}t} \proparp^{*} \ketplz \\
            &= (2\pi)^{-1/2} (2\Ep)^{-1} 
                   \int \dt'\, \theta(\pm t') \me^{\mp\mi\Ep(t-t')}
                   \int \dif p^{0}\, \me^{\mi p^{0}(t-t')} \ketplz \,.
    \end{split}
\end{equation*}
Change variables $t' \to t-t_{0}$ to get
\begin{equation} \label{eqn:N1a}
    \begin{split}
        \ketartp
            &= (2\pi)^{-1/2} (2\Ep)^{-1}
               \int dt_{0}\, \theta(\pm(t - t_{0}))
               \int \dif p^{0}\, \me^{\mi(p^{0} \mp \Ep)t_{0}} \ketplz \\
            &= \begin{cases}
                   (2\Ep)^{-1} \int_{-\infty}^{t} \dt_{0}\,
                                           \ketalz{t_{0}, \threep} \,, \\
                   (2\Ep)^{-1} \int_{t}^{+\infty} \dt_{0}\,
                                           \ketrlz{t_{0}, \threep} \,.
               \end{cases}
    \end{split}
\end{equation}

It is then straightforward to take the time limit $t \to \pm\infty$.
Note that
\begin{equation*}
    \begin{split}
        \int_{-\infty}^{+\infty} \dt_{0}\, \ketarlz{t_{0}, \threep}
            &= (2\pi)^{-1/2} \int \dif p^{0}\,
               \int dt_{0}\, \me^{\mi(p^{0} \mp \Ep)t_{0}} \ketplz \\
            &= (2\pi)^{-1/2} \int \dif p^{0}\,
               (2\pi) \delta(p^{0} \mp \Ep) \ketplz \\
            &= (2\pi)^{1/2} \ketarEplz \,.
    \end{split}
\end{equation*}
Therefore
\begin{equation} \label{eqn:N3}
    \ketarthreep \equiv \lim_{t \to \pm\infty} \ketartp
                     = (2\pi)^{1/2} (2\Ep)^{-1} \ketarEplz \,.
\end{equation}

Thus, a normal particle ($+$) or antiparticle ($-$) that has
3-momentum $\threep$ as $t \to \pm\infty$ is \emph{on-shell}, with
energy $\pm\Ep$. Such on-shell particles are unambiguously normal
particles or antiparticles, independent of choice of coordinate
system. (Note that these states are similar to the ``mass representation''
states of \refcite{horwitz73}.)

Note also that the factor of $\exp(\mp\mi\Ep t)$ in the definition of
$\ketartp$ (\eqn{eqn:N1}) is not arbitrary. Without this, a factor of
$\exp(\pm\mi\Ep t)$ would remain in \eqn{eqn:N1a}, making it
impossible to take the limit $t \to \pm\infty$.

\subsection{On-Shell Probability Interpretation}
\label{sect:free:probability}

Unfortunately, the states defined in \eqn{eqn:N3} are not normalizable 
using the usual inner product, since
\begin{equation*}
    \inner{\advret{\threepp}}{\advret{\threep}}
        = 2\pi (2\Ep)^{-2} \delta(0) \delta^{3}(\threepp - \threep)
\end{equation*}
is infinite. In \refcite{horwitz73}, this is handled by allowing the
mass $m$ to vary, even though the energy is fixed at
$\sqrt{\threep^{2} + m^{2}}$. Here we will take a different approach, 
noting that, from \eqn{eqn:J1},
\begin{equation*}
    \inner{p'; \lambda}{p; \lambda} = \delta^{4}(p' - p) \,.
\end{equation*}
Using this and \eqn{eqn:N3}, we clearly have
\begin{equation} \label{eqn:N6}
    \innerthreeparplz = (2\pi)^{1/2} (2\Ep)^{-1} \delta(\pm\Ep - p_{0}^{0})
                        \delta^{3}(\threep - \threep_{0}) \,.
\end{equation}
Moreover, from this equation and \eqn{eqn:N2},
\begin{equation} \label{eqn:N7}
    \begin{split}
        \inner{\advret{\threep}}{\advret{t_{0}, \threep_{0}{}}; \lambdaz}
            &= (2\pi)^{-1/2} \me^{\mp \mi\Ep t_{0}}
               \int \dif p_{0}^{0}\, \me^{\mi p_{0}^{0}t_{0}} 
               (2\pi)^{1/2} (2\Ep)^{-1} \delta(\pm\Ep - p_{0}^{0})
               \delta^{3}(\threep - \threep_{0}) \\
            &= (2\Ep)^{-1} \delta^{3}(\threep - \threep_{0}) \,,
    \end{split}
\end{equation}
for any value of $t_{0}$. This is essentially the basis for an
``induced'' inner product, in the sense of \refcites{hartle97,
halliwell01b}.

Let $\HilbH$ be the Hilbert space of the $\ketlz{x}$ and let
$\HilbH_{t}$ be the subspaces spanned by the $\ketlz{t,\threex}$, for
each $t$, forming a foliation of $\HilbH$. Now, from \eqn{eqn:N2}, it
is clear that the particle and antiparticle 3-momentum states
$\ketarlz{t,\threep}$ also each span $\HilbH_{t}$. In these
representations, states in $\HilbH_{t}$ have the form
\begin{equation} \label{eqn:N7a}
    \ketarlz{t,\psi} 
        = \intthree p\, \psi(\threep) \ketarlz{t,\threep} \,,
\end{equation}
for square-integrable functions $\psi(\threep)$. Conversely, it
follows from \eqn{eqn:N7} that a probability amplitude $\psi(\threep)$
is given by
\begin{equation} \label{eqn:N8}
    \psi(\threep) = (2\Ep)\inner{\advret{\threep}}
                                {\advret{t,\psi};\lambdaz} \,.
\end{equation}

Let $\HilbH'_{t}$ be the space of linear functions dual to
$\HilbH_{t}$. Via \eqn{eqn:N8}, the bra states $\braar{\threep}$ can
be considered to be members of $\HilbH'_{t}$, for all $t$. Indeed,
they span two common subspaces $\advret{\HilbH'}$ of the
$\HilbH'_{t}$, the states of which have the form
\begin{equation*}
    \braar{\psi}
        = \intthree p\, \psi(\threep)^{*} \braarthreep \,.
\end{equation*}
Now, define an inner product on the functions $\psi(\threep)$ such that
\begin{equation} \label{eqn:N8a}
    (\psi_{1}, \psi_{2}) 
        \equiv \inner{\advret{\psi_{1}{}}}
                     {\advret{t, \psi_{2}{}};\lambdaz} 
        = \int \frac{\dthree p}{2\Ep}
               \psi_{1}(\threep)^{*} \psi_{2}(\threep)\,,
\end{equation}
where the second equality follows from \eqn{eqn:N7}. Equipped with
this inner product, each $\HilbH_{t}$ is itself a Hilbert space of the
wave functions $\psi(\threep)$. Note that it is the states of the
\emph{dual} spaces $\advret{\HilbH'}$ that naturally satisfy the
on-shell constraint $\braar{\psi}\opH = 0$ (as suggested by, for
example, \refcite{giulini99}).

The operators $(2\Ep)\ketarlz{\tz,\threep}\braarthreep$ are
self-adjoint under the inner product given in \eqn{eqn:N8a}, in the
sense that the conjugate of
\begin{equation*}
    (2\Ep)\inner{\advret{\psi}}{\advret{\tz,\threep};\lambdaz}
    \braarthreep
        = \psi(\threep)^{*} \braarthreep
\end{equation*}
is
\begin{equation*}
    (2\Ep)\ketarlz{\tz,\threep}
    \inner{\advret{\threep}}{\advret{t,\psi};\lambdaz}
        = \ketarlz{\tz,\threep} \psi(\threep)
\end{equation*}
for that inner product. Further,
\begin{equation*}
    \intthree p\, (2\Ep)
        \inner{\advret{\psi_{1}{}}}{\advret{\tz,\threep};\lambdaz}
         \inner{\advret{\threep}}{\advret{t,\psi_{2}{}};\lambdaz}
         = \int \frac{\dthree p}{2\Ep}
                \psi_{1}(\threep)^{*} \psi_{2}(\threep)
         = \inner{\advret{\psi_{1}{}}}
                 {\advret{t, \psi_{2}{}};\lambdaz} \,,
\end{equation*}
which gives the effective resolution of the identity
\begin{equation} \label{eqn:N8b}
    \intthree p\, (2\Ep)\ketarlz{\tz,\threep}\braarthreep = 1 \,.
\end{equation}
In fact, such a resolution of the identity generally holds for
families of conjugate bra and ket states with a
\emph{bi-orthonormality relationship} such as \eqn{eqn:N7} (see
\refcite{akhiezer81} and App.\ A.8.1 of \refcite{muynk02}). We can,
therefore, take the operator $(2\Ep)\ketarlz{\tz,\threep}\braarthreep$
to represent the quantum proposition that an \emph{on-shell} particle
or antiparticle has the \emph{3-momentum} $\threep$. Then, with the
normalization
\begin{equation*}
    (\psi,\psi) 
        = \int \frac{\dthree p}{2\Ep} \sqr{\psi(\threep)} 
        = 1 \,,
\end{equation*}
$\sqr{\psi(\threep)}$ is the corresponding probability density in
3-momentum space.

Finally, consider that $\ketlz{t,\threex}$ is an eigenstate of the
3-position operator $\op{\threevec{X}}$, representing a particle
localized at the 3-position $\threex$ at time $t$. From \eqn{eqn:N8},
and using the inverse Fourier transform of \eqn{eqn:J1} with
\eqn{eqn:N3}, its 3-momentum wave function in $\HilbH_{t}$ is
\begin{equation} \label{eqn:N9}
    (2\Ep)\inner{\advret{\threep}}{t, \threex; \lambdaz}
        = (2\pi)^{-3/2} \me^{\mi(\pm\Ep t - \threep\cdot\threex)} \,.
\end{equation}
This is just a plane wave, and it is an eigenfunction of the operator
\begin{equation*}
         \me^{\pm\mi\Ep t}\, \mi\pderiv{}{\threep} \me^{\mp\mi\Ep t} \,,
\end{equation*}
which is the traditional momentum representation
$\mi\partial/\partial\threep$ of the 3-position operator
$\op{\threevec{X}}$, translated to time $t$. 

This result contrasts with the well-known result of Newton and Wigner
\cite{newton49}, who conclude that a localized particle wave function
satisfying the Klein-Gordon equation is an eigenfunction of
\begin{equation*}
    \mi \left( \pderiv{}{\threep} - \frac{\threep}{2\Ep^{2}} \right) \,,
\end{equation*}
which has an extra term over the expected
$\mi\partial/\partial\threep$. The key reason for this difference is
our use of the 3-momentum basis $\ketarlz{t, \threep}$. With the dual
basis $(2\Ep)\braar{\threep}$ from \eqn{eqn:N7}, this leads to the
relation given in \eqn{eqn:N8} and used in \eqn{eqn:N9}.

In contrast, the traditional formalism assumes that both bra and ket
states are on-shell. Instead of the time-dependent spaces
$\HilbH_{t}$, the spaces $\HilbH_{\pm}$ are used, with on-shell ket
basis states $\ketarthreep$ that are dual to the bra states
$\braarthreep$ under an inner product such that, instead of
\eqn{eqn:N7}, one has
\begin{equation*}
    (2\Epp)^{1/2}\inner{\advret{\threepp}}{\advret{\threep}}(2\Ep)^{1/2}
        = \delta^{3}(\threepp - \threep) \,,
\end{equation*}
where the factor of $(2\Ep)^{1/2}$ is introduced symmetrically on
dual bra and ket states in order provide an orthonormal basis. If we
were to use the traditional dual basis $(2\Ep)^{1/2}\braarthreep$,
instead of $(2\Ep)\braarthreep$, the wave function of
$\ketlz{t, \threex}$ would be
\begin{equation}  \label{eqn:N10}
    (2\Ep)^{1/2}\inner{\advret{\threep}}{t, \threex; \lambdaz}
        = (2\pi)^{-3/2} (2\Ep)^{-1/2} 
          \me^{\mi(\pm\Ep t - \threep\cdot\threex)} \,.
\end{equation}
At $t = 0$ this is exactly the Newton-Wigner wave function for a
localized particle \cite{newton49}.

Note that \eqn{eqn:N9} is effectively related to \eqn{eqn:N10} by a
scalar Foldy-Wouthuysen transformation \cite{foldy50, case54}. This
makes sense, since the Foldy-Wouthuysen transformation produces a
representation that separates positive and negative energy states
(particles and antiparticles) and gives a reasonable non-relativistic
limit.

Indeed, from \eqn{eqn:N2} we can easily see that the time evolution of
the 3-momentum states $\ketarlz{t, \threep}$ is given by
\begin{equation*}
    \me^{\mi\opP^{0}\Delta t}\ketarlz{t, \threep}
        = \me^{\pm\mi\Ep\Delta t}\ketarlz{t+\Delta t, \threep}
        = \me^{\pm\mi\op{H}_{FW}\Delta t}\ketarlz{t+\Delta t, \threep} \,,
\end{equation*}
where
\begin{equation*}
    \op{H}_{FW} = (\op{\threevec{P}}\cdot\op{\threevec{P}}
                   + m^{2})^{1/2}
\end{equation*}
is the scalar Foldy-Wouthuysen Hamiltonian and the $\opP^{\mu}$ are
the generators of spacetime translations. Define the operation of time
translation on the time-dependent states $\ketlz{t, \psi}$ so that
\begin{equation*}
    \ketarlz{t+\Delta t, \psi} 
        = \me^{\mi\opP^{0}\Delta t}\ketarlz{t,\psi} \,.
\end{equation*}
Substituting \eqn{eqn:N7a} then gives
\begin{equation*}
    \begin{split}
        \ketarlz{t+\Delta t, \psi}
            &= \intthree p\, \psi(t,\threep)
               \me^{\mi\opP^{0}\Delta t} \ketarlz{t,\threep} \\
            &= \intthree p\, \psi(t,\threep)
               \me^{\pm\mi\Ep\Delta t} \ketarlz{t+\Delta t, \threep} \\
            &= \intthree p\, \psi(t+\Delta t,\threep)
               \ketarlz{t+\Delta t, \threep} \,,
    \end{split}
\end{equation*}
where the time-dependence of the 3-momentum wave function has been
made explicit, with time evolution given by
\begin{equation*}
    \psi(t+\Delta t,\threep) = \me^{\pm\mi\Ep\Delta t}\psi(t,\threep) \,.
\end{equation*}
In the non-relativistic limit, for positive-energy particles, $\Ep
\approx m +\threep^{2}/2m$, and this time evolution
reduces to time evolution according to the usual non-relativistic
Hamiltonian (up to the momentum-independent phase factor
$\exp(\mi m\Delta t)$).

\subsection{Free Multiparticle States} \label{sect:free:multi}

The formalism introduced in the previous sections can be extended in a
straightforward way to a Fock space of non-interacting multiparticle
states. In order to allow for multiparticle states with different
types of particles, extend the position state of each individual
particle with a \emph{particle type index} $n$, such that
\begin{equation*}
    \inner{x',n';\lambda}{x,n;\lambda}
        = \delta^{n'}_{n}\delta^{4}(x'-x) \,.
\end{equation*}
Then, construct a basis for the Fock space of multiparticle states as
sym\-me\-trized products of $N$ single particle states:
\begin{equation} \label{eqn:X0}
    \ket{\xnliN}
        \equiv (N!)^{-1/2} \sum_{\text{perms }\Perm}
        \ket{\xni{\Perm 1};\lambda_{\Perm 1}} \cdots
        \ket{\xni{\Perm N};\lambda_{\Perm N}} \,,
\end{equation}
where the sum is over all permutations $\Perm$ of $1, 2, \ldots, N$.
(Since only scalar particles are being considered in the present work,
only Bose-Einstein statistics need be accounted for.)

Define multiparticle states $\ket{\xniN}$ as similarly sym\-me\-trized
products of $\ketx$ states. Then,
\begin{equation} \label{eqn:X}
    \inner{\xnpiN}{\seqN{\xnlzi}}
       = \sum_{\text{perms }\Perm}
               \prod_{i = 1}^{N} \delta^{n'_{\Perm i}}_{n_{i}}
               \propsym(x'_{\Perm i}-x_{i};m_{i}^{2}) \,,
\end{equation}
where $m_{i}$ is the mass of particles of type $n_{i}$. Note that the
use of the same parameter value $\lambdaz$ for the starting point of
each particle path is simply a matter of convenience, using the path
parameterization guage freedom to choose this value arbitrarily. The
intrinsic lengths of each particle path are still integrated over
\emph{separately} in $\ket{\xniN}$, which is important for obtaining
the proper particle propagator factors in \eqn{eqn:X}. Nevertheless,
by using $\lambdaz$ as a common starting parameter value, we can make
the small notational simplification of not repeating it multiple times
in $\ket{\seqN{\xnlzi}}$, defining, instead,
\begin{equation*}
    \ket{\xnlziN} \equiv \ket{\seqN{\xnlzi}} \,.
\end{equation*}

Following the same procedures as in \sect{sect:free:onshell} for each
particle in a multiparticle state, it is straightforward to construct
the multiparticle three momentum states $\ket{\tparniN}$ and
$\ket{\tparnlziN}$. Note that each particle may be \emph{either} a
normal particle ($+$) or an antiparticle ($-$). Then, to obtain
on-shell states we need to take $t_{i} \to +\infty$ in
$\ket{\tparniN}$ for particles, but $t_{i} \to -\infty$ for
antiparticles. This results in the multiparticle on-shell states
$\ket{\parniN}$.

Now, it can be seen that the $\ket{\parniN}$ states may not always be
particularly convenient, since they describe normal particles at $t =
+\infty$ and antiparticles at $t = -\infty$. For describing the
asymptotic state of outgoing particles from a scattering process, for
instance, we would like to take the limit for all particles and
antiparticles together as $t \to +\infty$.

To do this, we can take the viewpoint of considering antiparticles to
be positive energy particles traveling forwards in time, rather than
negative energy particles traveling backwards in time. Since both
particles and their antiparticles will then have positive energy, it
becomes necessary to explicitly label antiparticles with separate
(though related) types from their corresponding particles. Let $\na$
denote the type label for a normal particle type and $\nr$ denote the
corresponding antiparticle type.

For normal particles of type $\na$, position states are defined as
in \eqn{eqn:L1}:
\begin{equation*}
    \inner{x,\na}{\xz,\na;\lambdaz} = \thetaax\propa \,.
\end{equation*}
For antiparticles of type $\nr$, however, position states are now
defined such that
\begin{equation} \label{eqn:Y}
    \inner{x,\nr}{\xz,\nr;\lambdaz} = \thetaax\proprsym(\xz - x) \,.
\end{equation}
Note the reversal with respect to \eqn{eqn:L1} of $\xz$ and $x$ on 
the righthand side of this equation.

Using \eqn{eqn:L}, the Fourier transform of \eqn{eqn:Y} is
\begin{equation*}
    \begin{split}
        \intfour x\, \me^{-\mi p \cdot x} \theta(x^{0})
            \proprsym(-x)
            &= \intfour x\, \me^{-\mi p \cdot x}  
               \theta(x^{0}) \propasym(x^{0},-\threex) \\
            &= \intfour x\, \me^{\mi(p^{0}x^{0} + \threep\cdot\threex)} 
               \theta(x^{0}) \propasym(x) \\
            &= \propasym(p^{0},-\threep) \,,
    \end{split}
\end{equation*}
where $\propap$ is as given in \eqn{eqn:M}. From this we can see that
carrying through the derivation for antiparticle 3-momentum states
will, indeed, give positive energy states, but with reversed three
momentum:
\begin{equation*}
    \ket{t,\threep,\nr}
        = (2\Ep)^{-1}\int_{-\infty}^{t} \dt_{0}\,
          \ketlz{\tz,\threep,\nr} \,,
\end{equation*}
where
\begin{equation*}
    \ketlz{t_{0},\threep,\nr} = \ketlz{\adv{t_{0},-\threep},n} \,.
\end{equation*}
Further, taking the limit $t \to +\infty$ gives the on-shell states
\begin{equation*}
    \ket{\threep,\nr}
        \equiv \lim_{t \to +\infty} \ket{t,\threep,\nr}
             = (2\pi)^{1/2}(2\Ep)^{-1}\ketlz{+\Ep,-\threep} \,.
\end{equation*}

We can now reasonably construct Fock spaces $\HilbF_{t}$ with single
time, multiparticle basis states
\begin{equation*}
    \ketlz{t;\pnariN}
        \equiv \ketlz{t,\threep_{1},n_{1\pm};\ldots;
                      t,\threep_{N},n_{N\pm}} \,,
\end{equation*}
over all combinations of particle and antiparticle types. Similarly
defining
\begin{equation*}
    \ket{t;\pnariN} 
        \equiv \ket{t,\threep_{1},n_{1\pm};\ldots;
                    t,\threep_{N},n_{N\pm}} \,,
\end{equation*}
we can now take $t \to +\infty$ for particles and antiparticles alike 
to get the multiparticle on-shell states $\ket{\pnariN}$. The
corresponding bra states $\bra{\pnariN}$ then span a subspace of 
the dual space $\HilbF'_{t}$, for any $t$. Analogously to
the case for single particle states, this can be used to define a
Hilbert space of multiparticle probability amplitudes for each time
$t$.

Finally, since $\ket{\pnariN}$ now uniformly represents particles and
antiparticles in the $t \to +\infty$ limit, it can be used as the
asymptotically free state of outgoing particles from a scattering
process. The corresponding state for incoming particles is
$\ketlz{\pnariN} \equiv \lim_{t \to -\infty}\ketlz{t;\pnariN}$.

\subsection{Fields} \label{sect:free:fields}

Even though the theory presented here is essentially first-quantized,
it is still often convenient to introduce the formalism of creation
and annihilation fields on the Fock space of multi-particle states.
Specifically, define the creation field $\oppsit(x,n;\lambda)$ by
\begin{equation*}
    \oppsit(x,n;\lambda)\ket{\xnliN} 
        = \ket{x,n,\lambda;\xnliN} \,,
\end{equation*}
with the corresponding annihilation field $\oppsi(x,n;\lambda)$
having the commutation relation
\begin{equation*}
    [\oppsi(x',n';\lambda), \oppsit(x,n;\lambdaz)]
        = \delta^{n'}_{n}\propsym(x'-x;\lambda-\lambdaz) \,.
\end{equation*}
Further define
\begin{equation} \label{eqn:Z0}
    \oppsi(x,n) \equiv 
        \int_{\lambdaz}^{\infty} \dl\, \oppsi(x,n;\lambda) \,,
\end{equation}
so that
\begin{equation} \label{eqn:Z0a}
    [\oppsi(x',n'), \oppsit(x,n;\lambdaz)]
        = \delta^{n'}_{n}\propsym(x'-x) \,,
\end{equation}
which is consistent with the multi-particle inner product as given in
\eqn{eqn:X}.

Note the asymmetry in \eqn{eqn:Z0a}: $\oppsit(x,n;\lambdaz)$ is at the
reference value $\lambdaz$ of the path parameter (at the start of the
path), while in $\oppsi(x',n')$ the path parameter (at the end of the
path) is integrated over. This results from the fact that it is the
integrated position bra state $\bra{x',n'}$, created by
$\oppsi(x',n')$, that generates complete particle transition
amplitudes (as discussed at the end of \sect{sect:free:propagator}).
It is thus convenient to consider $\bra{x,n}$ to be ``dual'' to
$\ketlz{x,n}$, in a similar fashion to the states $\braarthreep$ and
$\ketlz{t,\advret{\threep}}$ in \sect{sect:free:onshell}, even though,
by \eqn{eqn:I2b}, the position states are not orthogonal.

In the field operator notation, this duality can be captured by
introducing a \emph{special adjoint} $\oppsi\dadj$ defined by
\begin{equation} \label{eqn:Z0b}
    \oppsi\dadj(x,n) = \oppsit(x,n;\lambdaz) \text{ and }
    \oppsi\dadj(x,n;\lambdaz) = \oppsit(x,n) \,.
\end{equation}
The commutation relation in \eqn{eqn:Z0a} then takes on the more 
symmetric form
\begin{equation*}
    [\oppsi(x',n'), \oppsi\dadj(x,n)]
        = \delta^{n'}_{n}\propsym(x'-x) \,.
\end{equation*}

We can also define field operators for explicit particle and
antiparticle types, as considered in \sect{sect:free:multi}. Define
the \emph{normal particle field} $\oppsi(x,\na)$ by
\begin{equation} \label{eqn:Z1}
    \oppsi(x,\na) \equiv \intfour \xz\,
                         \propa \oppsi(\xz,\na;\lambdaz) \,,
\end{equation}
giving the commutation rule
\begin{equation} \label{eqn:Z2}
    [\oppsi(x',\na), \oppsit(x,\na;\lambdaz)]
        = [\oppsi(x',\na), \oppsi\dadj(x,\na)]
        = \propasym(x'-x) \,.
\end{equation}
Substituting \eqn{eqn:L} into \eqn{eqn:Z1} gives the familiar
expression
\begin{equation*}
    \oppsi(x,\na) = (2\pi)^{-3/2} \intthree p\, 
        \me^{\mi(-\Ep x^{0} + \threep\cdot\threex)}
        \op{a}(\threep,\na) \,,
\end{equation*}
where
\begin{equation*}
    \op{a}(\threep,\na) \equiv
        (2\pi)^{-3/2} (2\Ep)^{-1} \intfour \xz\,
            \me^{\mi(\Ep \xz^{0} - \threep\cdot\threex_{0})}
            \oppsi(\xz,\na;\lambdaz)
\end{equation*}
is the \emph{on-shell particle 3-momentum field}.

For antiparticles, reverse the roles of the antiparticle creation and
annihilation operators relative to increasing-$\lambda$ propagation as
defined for the normal particle type. Define the antiparticle
\emph{creation} field analogously to \eqn{eqn:Z1} for the
corresponding normal particle \emph{annihilation} field:
\begin{equation*}
    \oppsit(x,\nr) 
        \equiv \intfour \xz\,
               \propr \oppsit(\xz,\nr;\lambdaz) \,.
\end{equation*}
Now, $\propr^{*} = \proprsym(\xz-x)$ (see \eqn{eqn:L}). Therefore,
\begin{equation} \label{eqn:Z3}
    \oppsi(x,\nr) = \intfour \xz\,
        \proprsym(\xz-x) \oppsi(\xz,\nr;\lambdaz) \,,
\end{equation}
giving the commutation rule (note the switching of $x'$ and $x$ 
on the right, relative to \eqn{eqn:Z2})
\begin{equation*}
    [\oppsi(x',\nr), \oppsit(x,\nr;\lambdaz)]
        = [\oppsi(x',\nr), \oppsi\dadj(x,\nr)]
        =  \proprsym(x-x') \,.
\end{equation*}
Substituting \eqn{eqn:L} into \eqn{eqn:Z3} and changing variables
$\threep \to -\threep$ then gives
\begin{equation*}
    \oppsi(x,\nr) = (2\pi)^{-3/2} \intthree p\, 
        \me^{\mi(-\Ep x^{0} + \threep\cdot\threex)}
        \op{a}(\threep,\nr) \,,
\end{equation*}
where
\begin{equation*}
    \op{a}(\threep,\nr) \equiv
        (2\pi)^{-3/2} (2\Ep)^{-1} \intfour \xz\,
            \me^{\mi(\Ep \xz^{0} - \threep\cdot\threex_{0})}
            \oppsi(\xz,\nr;\lambdaz)
\end{equation*}
is the \emph{on-shell antiparticle 3-momentum field}.

\section{Interacting Particles} \label{sect:interacting}

In conventional second-quantized quantum field theory, interactions
are introduced via the Lagrangian density into the Hamiltonian used
to propagate the fields. The very conception of interacting particles
and their paths then only arises at all as a result of the
perturbative expansion of the Hamiltonian. Such an approach is thus 
not very natural for a foundational formalism based on spacetime paths.

Now, the actual traditional motivation for introducing fields in the
first place is largely a heuristic response to the well known
difficulties with negative energies and probabilities in relativistic
quantum mechanics. However, as we have seen in \sect{sect:free}, these
difficulties can also be handled in the context of the spacetime path
formalism. Further, the spacetime path approach can very directly
accommodate the creation and destruction of particles, as required in
a relativistic theory. One simply considers particle paths with a
finite length: a particle is created at the start of its path and
destroyed at the end.

Taking this path viewpoint, an \emph{interaction vertex} can then
simply be considered as an event at which a set of particle paths all
end together and another set of particle paths all begin. An
\emph{interaction graph} is a set of interaction vertices connected by
particle paths. For a collection of interacting particles, it is
essentially such graphs that act as the fundamental building blocks of
the system state, rather than the individual particle paths
themselves. 

The natural spacetime path approach for interactions is therefore
first quantized rather than second quantized. As we will see in this
section, the first-quantized spacetime path formalism can duplicate
the basic results of perturbative quantum field theory for Feynman
diagrams and scattering. It is also consistent with the typically
first-quantized geometric approach used in string theory
\cite{green87}.

Of course, taking a first-quantized formalism as foundational requires
that issues of consistency and convergence that appear in traditional
perturbation theory be addressed directly, without recourse to a
posited non-perturbative solution. We will return to this point at the
end of \sect{sect:interacting:renormalization}, though a full
discussion is beyond the scope of the present paper.

\subsection{Interactions} \label{sect:interacting:interactions}

Since incoming particles are destroyed at an interaction vertex, and
outgoing particles are created, the vertex can be represented by an
operator constructed as an appropriate product of the creation and
annihilation operator fields introduced in \sect{sect:free:fields}.
Note that ``incoming'' and ``outgoing'' are used here in the sense of
the path evolution parameter $\lambda$, not time. That is, we are not
separately considering particles and antiparticles at this point.

\begin{postulate}
    An interaction vertex, possibly occurring at any position in
    spacetime, with some number $a$ of incoming particles and some
    number $b$ of outgoing particles, is represented by the operator
    \begin{equation} \label{eqn:Y0}
        -\mi\opV \equiv h \intfour x\,
                               \prod_{i = 1}^{a} 
                                   \oppsi\dadj(x,n'_{i})
                               \prod_{j = 1}^{b} \oppsi(x,n_{j}) \,,
    \end{equation}
    where the coefficient $h$ represents the relative probability
    amplitude of the interaction and $\oppsi\dadj$ is the special 
    adjoint defined in \eqn{eqn:Z0b}.
\end{postulate}

The probability amplitude for a transition from an initial state
$\ket{\xnlziN}$ to a final state $\ket{\xnpiNp}$, with
a single intermediate interaction, is then
\begin{multline*}
    G_{1}(\xnpiNp | \xniN) \\
        = \bra{\xnpiNp} (-\mi)\opV \ket{\xnlziN} \,.
\end{multline*}
This is essentially the amplitude for a first-order \emph{Wick
diagram} \cite{ticciati99}. That is, it is equivalent to the
first-order terms in the Wick expansion of the Dyson series in
conventional quantum field theory (including all permutations that may
result from crossing symmetries if any of the incoming or outgoing
particles in the interaction are of the same type).

The probability amplitude corresponding to multiple intermediate 
interactions can then be obtained by repeated applications of $\opV$. 
Thus, the amplitude for $m$ interactions is
\begin{multline*}
    G_{m}(\xnpiNp | \xniN) \\
        = \bra{\xnpiNp} \frac{(-\mi)^{m}}{m!}\opV^{m}
          \ket{\xnlziN} \,,
\end{multline*}
where the $(m!)^{-1}$ factor accounts for all possible permutations of
the $m$ identical factors of $\opV$. The \emph{complete} interacting
transition amplitude, with any number of intermediate interactions, is
then
\begin{equation} \label{eqn:Y1}
    \begin{split}
        G(\xnpiNp | \xniN)
            &= \sum_{m=0}^{\infty} G_{m}(\xnpiN | \xniN) \\
            &= \bra{\xnpiN} \opG \ket{\xnlziN} \,,
    \end{split}
\end{equation}
where
\begin{equation} \label{eqn:Y1a}
    \opG \equiv \sum_{m=0}^{\infty} \frac{(-\mi)^{m}}{m!}\opV^{m}
              = \me^{-i\opV} \,.
\end{equation}
Extend the operation of the special adjoint in the natural way
to sums and products. Then it is clear, at least formally, that $\opG$
is unitary relative to this adjoint (that is, $\opG\dadj\opG =
\opG\opG\dadj = 1$), so long as $\opV$ is self-adjoint relative to it
(that is, $\opV\dadj = \opV$).

From \eqn{eqn:Y0}, there are two consequences to $\opV$ being
self-adjoint. First $\mi h = g$ must be real. Second, the interaction
must involve the same number of incoming and outgoing particles, of
the same types. This second consequence is a result of assuming so far
that there is only one possible type of interaction. The formalism can
be easily extended to allow for multiple types of interactions by
adding additional terms to the definition of $\opV$. In this case,
only the overall operator $\opV$ needs to be self-adjoint, not the
individual interaction terms.

For example, consider the case of a three-particle interaction of the
form $\oppsi\dadj(x,n_{A})\oppsi(x,n_{B})\oppsi(x,n_{A})$. Then, for
the overall interaction operator $\opV$ to be self-adjoint, there must
also be a conjugate interaction term
$\oppsi\dadj(x,n_{A})\oppsi\dadj(x,n_{B})\oppsi(x,n_{A})$. That is,
\begin{equation*}
    \opV 
        = g \intfour x\,
              [ \oppsi\dadj(x,n_{A})\oppsi(x,n_{B})
                \oppsi(x,n_{A})
              + \oppsi\dadj(x,n_{A})\oppsi\dadj(x,n_{B})
                \oppsi(x,n_{A}) ] \,.
\end{equation*}
This corresponds to the case of the particle of type $B$ being
indistinguishable from its antiparticle. Defining the self-adjoint
effective field
\begin{equation*}
    \oppsi'(x,n_{B}) 
        \equiv \oppsi(x,n_{B}) + \oppsi\dadj(x,n_{B})
\end{equation*}
then allows $\opV$ to be put back into the form of a single type of 
interaction:
\begin{equation} \label{eqn:Y2}
    \opV 
        = g \intfour x\,
              \oppsi\dadj(x,n_{A})\oppsi'(x,n_{B})\oppsi(x,n_{A}) \,.
\end{equation}

An alternate interpretation of a self-adjoint interaction vertex is to
pair up incoming and outgoing particles of the same type and consider
them to be the \emph{same} particle before and after the interaction.
For example, an interaction of the form given in \eqn{eqn:Y2} would be
considered to represent a \emph{single} particle of type $A$
interacting with a self-adjoint particle of type $B$.

This viewpoint can be seen more explicitly by considering a
first-order interaction matrix element and using \eqn{eqn:I2a} to
expand the $A$-particle propagators:
\begin{equation*}
    \begin{split}
        &\bra{x_{A},n_{A};x_{B},n_{B}}\opV\ketlz{\xz,n_{A}} \\
        &\qquad = g \intfour x\,
                  \propsym_{A}(x_{A}-x)\propsym_{B}(x_{B}-x)
                  \propsym_{A}(x-\xz) \\
        &\qquad = g \intfour x\,
                  \propsym_{B}(x_{B}-x)
                  \int_{\lambdaz}^{\infty} \dl \,
                  \int_{\lambdaz}^{\infty} \dl' \,
                  \kersym_{A}(x_{A}-x;\lambda'-\lambdaz)
                  \kersym_{A}(x-\xz;\lambda-\lambdaz) \\
        &\qquad = g \intfour x\,
                  \propsym_{B}(x_{B}-x)
                  \int_{\lambdaz}^{\infty} \dl \,
                  \int_{\lambda}^{\infty} \dl' \,
                  \kersym_{A}(x_{A}-x;\lambda'-\lambda)
                  \kersym_{A}(x-\xz;\lambda-\lambdaz) \,.
    \end{split}
\end{equation*}
Substituting for the $A$-particle kernels from \eqn{eqn:B}, the path
integral for the first kernel ends at the same point $x$ as the path
integral for the second kernel begins. Therefore, the two path
integrals can be combined into a single path integral, with the
constraint that the paths always pass through the intermediate point
$x$:
\begin{equation*}
    \begin{split}
        &\bra{x_{A},n_{A};x_{B},n_{B}}\opV\ketlz{\xz,n_{A}} \\
        &\qquad = \begin{aligned}[t]
                      g \intfour x\,
                      \propsym_{B}(x_{B}-x)
                      \int_{\lambdaz}^{\infty} \dl \,
                      \int_{\lambda}^{\infty} \dl' \,
                      \eta \intDfour q\, 
                          \delta^{4}(q(\lambda') - x_{A})
                          &\delta^{4}(q(\lambda) - x) \\
                          &\delta^{4}(q(\lambdaz) - \xz)
                          \me^{\mi S_{A}[\qdot]}
                  \end{aligned} \\
        &\qquad = g \int_{\lambdaz}^{\infty} \dl' \,
                  \int_{\lambdaz}^{\lambda'} \dl \,
                  \eta \intDfour q\,                   
                      \delta^{4}(q(\lambda') - x_{A})
                      \delta^{4}(q(\lambdaz) - \xz)
                      \me^{\mi S_{A}[\qdot]}
                      \propsym_{B}(x_{B}-q(\lambda)) \,.
    \end{split}
\end{equation*}
This form clearly reflects the viewpoint of a single $A$-particle
interacting with a $B$-particle at the point $q(\lambda)$ in its path.

Now consider a second-order interaction in which the incoming and
outgoing particles are all $A$-particles:
\begin{multline} \label{eqn:Y2a}
    \bra{x'_{1},n_{A};x'_{2},n_{A}}\frac{1}{2}\opV^{2}
    \ketlz{x_{1},n_{A};x_{2},n_{A}} \\
        = \frac{g^{2}}{2} \intfour y_{1}\, \intfour y_{2}\,
          [ \propsym_{A}(x'_{1}-y_{2})
            \propsym_{A}(y_{2}-y_{1})
            \propsym_{A}(y_{1}-x_{1})
            \propsym_{B}(y_{2}-y_{1}) 
            \propsym_{A}(x'_{2}-x_{2})\\
          + \propsym_{A}(x'_{1}-y_{1})
            \propsym_{A}(y_{1}-x_{1})
            \propsym_{B}(y_{2}-y_{1})
            \propsym_{A}(x'_{2}-y_{2})
            \propsym_{A}(y_{2}-x_{2})
          + \cdots ] \,,
\end{multline}
where the additional terms not shown are the result of position
interchanges from the terms given. The first term shown in
\eqn{eqn:Y2a} reflects a self-interaction of one $A$ particle via the
$B$ particle, with the second $A$ particle propagating freely. The
self-interaction factor can be given the path integral representation
\begin{multline} \label{eqn:Y2b}
    \intfour y_{1}\, \intfour y_{2}\,
    \propsym_{A}(x'_{1}-y_{2})
    \propsym_{A}(y_{2}-y_{1})
    \propsym_{A}(y_{1}-x_{1})
    \propsym_{B}(y_{2}-y_{1}) \\
        = \int_{\lambdaz}^{\infty} \dl'
          \int_{\lambdaz}^{\lambda'} \dl_{2}
          \int_{\lambdaz}^{\lambda_{2}} \dl_{1} \,
          \eta \intDfour q\,                   
              \delta^{4}(q(\lambda') - x'_{1})
              \delta^{4}(q(\lambdaz) - x_{1})
              \me^{\mi S_{A}[\qdot]}
              \propsym_{B}(q(\lambda_{2})-q(\lambda_{1})) \,,
\end{multline}
reflecting an $A$ particle interacting with the $B$ particle at points
$\lambda_{1}$ and $\lambda_{2}$. The second term shown in
\eqn{eqn:Y2a} reflects an interaction of two $A$ particles via a $B$
particle. It can be given the path integral representation
\begin{multline*}
    \intfour y_{1}\, \intfour y_{2}\,
    \propsym_{A}(x'_{1}-y_{1})
    \propsym_{A}(y_{1}-x_{1})
    \propsym_{A}(x'_{2}-y_{2})
    \propsym_{A}(y_{2}-x_{2})
    \propsym_{B}(y_{2}-y_{1}) \\
        = \int_{\lambdaz}^{\infty} \dl'_{2} \,
          \int_{\lambdaz}^{\infty} \dl'_{1} \,
          \int_{\lambdaz}^{\lambda'_{2}} \dl_{2}\,
          \int_{\lambdaz}^{\lambda'_{1}} \dl_{1} \,
          \eta^{2} \intDfour q_{2}\, 
              \delta^{4}(q_{2}(\lambda'_{2})-x'_{2})
              \delta^{4}(q_{2}(\lambdaz) - x_{2})
              \me^{\mi S_{A}[\qdot_{2}]} \\
              \times
              \intDfour q_{1}\,
                  \delta^{4}(q_{1}(\lambda'_{1}) - x'_{1})
                  \delta^{4}(q_{1}(\lambdaz) - x_{1})
                  \me^{\mi S_{A}[\qdot_{1}]}
                  \propsym_{B}
                      (q_{2}(\lambda_{2})-q_{1}(\lambda_{1})) \,,
\end{multline*}
showing the $B$ particle propagating from the point at $\lambda_{1}$
on the path of the first $A$ particle to the point at $\lambda_{2}$ on
the path of the second $A$ particle. If the $B$ particle is taken to
be a photon, then Barut and Duru have shown that expansions of just
the form given above can be obtained from a general path integral
formulation of quantum electrodynamics \cite{barut89} (see also the
similar result obtained in \refcite{horwitz82} using a parametrized
perturbation series approach).

\subsection{Feynman Diagrams} \label{sect:interacting:feynman}

Computing a scattering amplitude requires moving from the Wick diagram
formulation of \eqn{eqn:Y1} to a Feynman diagram formulation. To do
this, replace the initial and final states in \eqn{eqn:Y1} with
on-shell multiparticle momentum states $\ket{\tparnlziN}$ and
$\ketlz{\parnpiN}$ (note that these are the on-shell
multiparticle states defined in \sect{sect:free:multi}, with
antiparticles propagating \emph{backwards} in time, \emph{not} the
single-time states defined at the end of that section):
\begin{multline} \label{eqn:Y3}
    G(\parnpiN | \parniN) \\
        \equiv \left[ \prod_{i=1}^{N'} 2\E{\threepp_{i}}
                          \prod_{i=1}^{N} 2\E{\threep_{i}}
                    \right]^{1/2}
               \bra{\parnpiN} \opG \ket{\tparnlziN} \,.
\end{multline}
The $2\Ep$ factors are required by the resolution of the identity for
these multi-particle states, generalizing the single particle case of
\eqn{eqn:N8b}:
\begin{multline} \label{eqn:Y4}
    \sum_{N = 0}^{\infty}\, \sum_{n_{i}\pm}
    \int \dthree p_{1} \cdots \dthree p_{N}\,
        \left[ \prod_{i=1}^{N} 2\E{\threep_{i}} \right] \\
        \times \ket{\tparnlziN}\bra{\parniN}
        = 1 \,,
\end{multline}
where the summation over the $n_{i}\pm$ is over all particle types 
\emph{and} particle/antiparticle states.

Note that use of the on-shell states in \eqn{eqn:Y3} requires
specifically identifying external lines as particles and
antiparticles. For each initial and final particle, $+$ is chosen if
it is a normal particle and $-$ if it is an antiparticle. The result
is a sum of Feynman diagrams, including all possible permutations of
interaction vertices and crossing symmetries. The inner products of
the on-shell states for individual initial and final particles with
the off-shell states for interaction vertices give the proper factors
for the external lines of a Feynman diagram.

For a final particle, the on-shell state $\braa{\threepp}$ is obtained
in the limit $t' \to +\infty$. Such a particle is thus an
\emph{outgoing} particle from the scattering process. If the external
line for this particle starts at an interaction vertex $x$, then the
line contributes an appropriate factor
\begin{equation*}
    (2\Epp)^{1/2} \inner{\adv{\threepp}}{x;\lambdaz}
        = (2\pi)^{-3/2} (2\Epp)^{-1/2}
          \me^{\mi(+\Epp x^{0} - \threepp \cdot \threex)} \,.
\end{equation*}

For a final antiparticle, however, the on-shell state
$\brar{\threepp}$ is obtained in the limit $t' \to -\infty$. This
means that the antiparticle is \emph{incoming} to the scattering
process, even though it derives from a final vertex, reflecting the
time-reversal of antiparticle paths. If the external line for this
antiparticle starts at an interaction vertex $x$, then the line
contributes the factor
\begin{equation*}
    (2\Epp)^{1/2} \inner{\ret{\threepp}}{x;\lambdaz} 
        = (2\pi)^{-3/2} (2\Epp)^{-1/2}
          \me^{\mi(-\Epp x^{0} + \threepp \cdot \threex)} \,.
\end{equation*}

Next, consider an initial particle on an external line ending at an
interaction vertex $x'$, The factor for this line is (assuming 
$x^{\prime 0} > t$)
\begin{equation*}
    (2\Ep)^{1/2} \inner{x'}{\adv{t,\threep};\lambdaz}
        = (2\pi)^{-3/2} (2\Ep)^{-1/2}
        \me^{\mi(-\Ep x^{\prime 0} + \threep \cdot \threexp)} \,.
\end{equation*}
Note that this expression is independent of $t$, so we can take $t \to
-\infty$ and treat the particle as \emph{incoming}. For an initial
antiparticle, the corresponding factor is (assuming $x^{\prime 0} <
t$)
\begin{equation*}
    (2\Ep)^{1/2} \inner{x'}{\ret{t,\threep};\lambdaz}
        = (2\pi)^{-3/2} (2\Ep)^{-1/2}
          \me^{\mi(+\Ep x^{\prime 0} - \threep \cdot \threexp)} \,.
\end{equation*}
Taking $t \to +\infty$, this represents the factor for an antiparticle
that is \emph{outgoing}.

If a particle or antiparticle both starts at an initial vertex $x$ and
ends at a final vertex $x'$, then, by \eqn{eqn:N7},
\begin{equation*}
    (2\Epp 2\Ep)^{1/2} \inner{\advret{\threepp}}
                             {\advret{t, \threep};\lambdaz}
        = \delta^{3}(\threepp - \threep) \,.
\end{equation*}
Finally, particles that start and end on interaction vertices (i.e., 
internal edges) are ``virtual'' particles propagating between
interactions, retaining the full Feynman propagator factor
$\propsym(x'-x)$.

Thus, the effect of \eqn{eqn:Y3} is to remove the propagator factors
from the external lines of the summed Feynman diagrams, retaining them
on internal edges. Since, in the position representation, $G$ is
essentially a sum of Green's functions $G_{m}$, this procedure is
effectively equivalent to the usual LSZ reduction of the Green's
functions \cite{lsz55, ticciati99, weinberg95}.

\subsection{Scattering} \label{sect:interacting:scattering}

The formulation of \eqn{eqn:Y3} is still not that of the usual
scattering matrix, since the initial state involves incoming particles
but outgoing antiparticles, and vice versa for the final state. To
construct the usual scattering matrix, it is necessary to have
multiparticle states that involve either all incoming particles and
antiparticles (that is, they are composed of individual asymptotic
particle states that are all consistently for $t \to -\infty$) or all
outgoing particles and antiparticles (with individual asymptotic
states all for $t \to +\infty$). These are the states
$\ketlz{\pnariN}$ and $\ket{\pnariN}$ defined at the end of
\sect{sect:free:multi}.

Reorganizing the scattering amplitude of \eqn{eqn:Y3} in terms of
these asymptotic states gives the more usual form using the scattering
operator $\opS$. Showing explicitly the asymptotic time limit used for
each particle:
\begin{multline} \label{eqn:Z3a}
    \bra{+\infty,\adv{\threepp},n';\ldots;
         -\infty,\ret{\bar{\threep}'},\bar{n}';\ldots} \opG
    \ketlz{{-\infty},\adv{\threep},n;\ldots;
           +\infty,\ret{\bar{\threep}},\bar{n};\ldots} \\
    = \bra{\threepp,\na';\ldots;
           \bar{\threep},\ret{\bar{n}};\ldots} \opS
      \ketlz{\threep,\na;\ldots;
             \bar{\threep}',\ret{\bar{n}}';\ldots} \,.
\end{multline}
Using the resolution of the identity
\begin{multline} \label{eqn:Z3b}
    \sum_{N = 0}^{\infty}\, \sum_{\advret{n_{i}{}}}
    \int \dthree p_{1} \cdots \dthree p_{N}\,
        \left[ \prod_{i=1}^{N} 2\E{\threep_{i}} \right] \\
        \times \ket{\pnarlziN}\bra{\pnariN}
        = 1 \,,
\end{multline}
expand the state $\opS\ket{\pnarlziN}$ as
\begin{multline*}
    \opS\ket{\pnarlziN} \\
        = \sum_{N' = 0}^{\infty}\, \sum_{\advret{n_{i}{}}}
          \int \dthree p'_{1} \cdots \dthree p'_{N'}\,
          \left[ \prod_{i=1}^{N'} 2\E{\threepp_{i}} \right]
           \ket{\pnparlziN} \\
           \times \bra{\pnpariN}\opS\ket{\pnarlziN} \,.
\end{multline*}
This shows how $\opS\ket{\pnarlziN}$ is a superposition of 
possible out states, with the square of the scattering amplitude, 
\eqn{eqn:Z3a}, giving the probability of a particular out state for a 
particular in state.

Next, use \eqns{eqn:Z2} and \eqref{eqn:Z3} in \eqn{eqn:K1} to write the
propagator as
\begin{equation} \label{eqn:Z4}
    \prop = 
        \thetaax [\oppsi(x,\na), \oppsi\dadj(\xz,\na)] +
        \thetarx [\oppsi(\xz,\nr), \oppsi\dadj(x,\nr)] \,.
\end{equation}
Then, reversing the usual derivation for Feynman diagrams (see, for
example, \refcite{weinberg95}) gives the Dyson series expansion
\begin{equation} \label{eqn:Z5}
    \opS = 1 + \sum_{n=1}^{\infty} \frac{(-\mi)^{n}}{n!}
                   \int_{-\infty}^{\infty} \dt_{1}\,
                   \int_{-\infty}^{t_{1}} \dt_{2}\, \cdots
                   \int_{-\infty}^{t_{n-1}} \dt_{n}\,
                   \opV(t_{1})\opV(t_{2}) \cdots \opV(t_{n})
\end{equation}
in terms of the \emph{time-dependent interaction operator}
\begin{equation*}
    \opV(t)
        \equiv g \intthree x\,
               \prod_{i = 1}^{a} \opPsi\dadj(t,\threex,n'_{i})
               \prod_{j = 1}^{b} \opPsi(t,\threex,n_{j}) \,,
\end{equation*}
where
\begin{equation} \label{eqn:Z5a}
    \opPsi(x,n) \equiv \oppsi(x,\na) + \oppsi\dadj(x,\nr) \,.
\end{equation}

Since $\opV(t)$ represents an interaction with the same number of
incoming and outgoing particles, of the same types, as $\opV$, the
self-adjointness of $\opV$ implies the self-adjointness of $\opV(t)$,
from which it can be shown that $\opS$ is unitary. The case of a
self-adjoint effective field $\oppsi'(x,n)$ in $\opV$ (as discussed at
the end of \sect{sect:interacting:interactions}) corresponds to the
requirement of self-adjointness for $\opPsi(x,n)$. As can be seen from
\eqn{eqn:Z5a}, this requirement implies that particles of type $n$ are
indistinguishable from their (path-reversed) antiparticles (indeed, a
working definition of ``indistinguishable'' in this sense might very
well be ``cannot be distinguished by any interaction'').

\subsection{Regularization and Renormalization}
\label{sect:interacting:renormalization}

Of course, the development given in the previous subsections is
actually only formal, because of the usual problems with divergence of
the series in \eqn{eqn:Y1}. As in conventional field theory, it is
necessary to regularize infinite integrals and renormalize the
resulting amplitudes. For a first-quantized approach, though, these
problems seem particularly severe, since \eqn{eqn:Y1} is taken as the
fundamental definition for the interacting amplitude, rather than as a
perturbation expansion.

Fortunately, there is a relatively straightforward way to approach
regularization within the context of a spacetime path approach,
inspired by the work of Frastai and Horwitz \cite{frastai95} (see also
\refcites{land97,land03} for a similar approach in the context of
off-shell electrodynamics). This is to make the interaction coupling
dependent on the intrinsic path length. This can be naturally
introduced into the spacetime path formalism by making a choice for
the weight function $f(T)$ introduced in \sect{sect:free:propagator}
different than $f(T) = 1$.

To see this, consider that replacing the field operator $\oppsi(x,n)$
defined in \eqn{eqn:Z0} with
\begin{equation*}
    \oppsi_{f}(x,n) \equiv \int_{\lambdaz}^{\infty} \dl\, 
                               f(\lambda-\lambdaz)\oppsi(x,n;\lambda)
\end{equation*}
gives the commutation relation
\begin{equation*}
    [\oppsi_{f}(x',n), \oppsit(x,n;\lambdaz)]
        = \int_{0}^{\infty} \dif T\, f(T) \kersym(x'-x;T) \,,
\end{equation*}
resulting in a propagator including the weight factor $f(T)$. Using
this new field operator for, say, particles of type $n_{A}$ in the
interaction vertex operator given in \eqn{eqn:Y2} produces the desired
path-length-dependent coupling:
\begin{equation} \label{eqn:Z6}
    \opV 
        = g \intfour x\, \int_{\lambdaz}^{\infty} \dl\,
              f(\lambda-\lambdaz)
              \oppsit(x,n_{A};\lambdaz) \oppsi(x,n_{A};\lambda)
              \oppsi'(x,n_{B}) \,.
\end{equation}
For the purposes of the present section, an appropriate choice for
$f(\lambda - \lambdaz)$ is the Gaussian
\begin{equation*}
    f(\lambda-\lambdaz) 
        = \me^{-(\lambda-\lambdaz)^{2}/2\Delta\lambda^{2}} \,,
\end{equation*}
where $\Delta\lambda$ is a \emph{correlation length}. For
$\Delta\lambda \to \infty$, $f(\lambda-\lambdaz) \to 1$, and
\eqn{eqn:Z6} reduces to \eqn{eqn:Y2}.

Now, consider again the self-interaction term from \eqn{eqn:Y2a}.
Using the interaction vertex operator from \eqn{eqn:Z6}, this becomes
\begin{equation*}
    \propsym_{A}(p) \intfour p'\,
    \int_{\lambdaz}^{\infty} \dl_{1}\, 
    \int_{\lambdaz}^{\infty} \dl_{2}\,
    f(\lambda_{2}-\lambdaz) \kersym_{A}(p';\lambda_{2}-\lambda_{0})
    f(\lambda_{1}-\lambdaz) \propsym_{B}(p-p')
    \kersym_{A}(p;\lambda_{1}-\lambda_{0})
\end{equation*}
For simplicity, the momentum representation has been used here, in
which
\begin{equation*}
    \kersym_{A}(p;\lambda-\lambdaz)
        \equiv \me^{-\mi(\lambda-\lambdaz)(p^{2}+m_{A}^{2})}
\end{equation*}
and
\begin{equation*}
    \propsym_{A}(p) \equiv \int_{0}^{\infty} \dif T\, 
                               \kersym_{A}(p;T)
                         = -\mi(p^{2}+m_{A}^{2}-\mi\varepsilon)
\end{equation*}
(and similarly for $\propsym_{B}$). The propagator from $\lambdaz$ to
$\lambda_{1}$ is not divergent, so we can let $f(\lambda_{1}-\lambdaz)
\to 1$, giving $\propsym_{A}(p) T'(p) \propsym_{A}(p)$, where
\begin{equation} \label{eqn:Z8}
    T'(p) \equiv \intfour p'\,
                     \int_{\lambdaz}^{\infty} \dl\, 
                     f(\lambda-\lambdaz)\kersym_{A}(p';\lambda-\lambdaz)
                     \propsym_{B}(p-p') \,.
\end{equation}

Inserting \eqn{eqn:I3} into into \eqn{eqn:Z8} gives
\begin{equation} \label{eqn:Z10}
    T'(p) = \int \dif m^{2}\, T(p;m^{2}) F(m^{2}) \,,
\end{equation}
where
\begin{equation*}
    T(p;m^{2}) \equiv \intfour p'\, \propsym(p';m^{2})
                                    \propsym_{B}(p-p') \,,
\end{equation*}
is the unregulated self-interaction amplitude (without the external
legs), with
\begin{equation}
    \propsym(p;m^{2}) \equiv \int_{0}^{\infty} \dl'\, 
                             \me^{-\mi\lambda'(p^{2}+m^{2})} \\
                      =      -i(p^{2}+m^{2}-i\varepsilon) \,,
\end{equation}
and
\begin{equation*}
    F(m^{2}) \equiv (2\pi)^{-1} \int_{0}^{\infty} \dl\,
                    \me^{\mi\lambda(m^{2}-m_{A}^{2})}
                    f(\lambda) \,.
\end{equation*}
The unregulated quantity $T(p;m_{A}^{2})$ is divergent. However,
\eqn{eqn:Z10} is exactly the Pauli-Villars regularization prescription
in continuous form \cite{pauli49}. Adjust the Fourier transform of the
coefficients $F(m^{2})$ so that
\begin{equation*}
    \tilde{F}(\lambda) =
        \begin{cases}
            f(\lambda) \me^{-\mi\lambda m_{A}^{2}}, 
                &\text{if $\lambda > \delta$;} \\
            0,  &\text{if $\lambda \leq \delta$.}
        \end{cases}
\end{equation*}
This then meets the Pauli-Villars conditions in Fourier space for
cancelation of singularities \cite{schwinger51}: $\tilde{F}(0) = 0$
and $\tilde{F}'(0) = 0$. For $\Delta\lambda \to \infty$ and $\delta
\to 0$, $T'(p)$ reduces to the unregulated quantity $T(p;m_{A}^{2})$.
(For further discussion, see \refcite{frastai95}. In \refcites{land97,
land03}, a similar result is obtained for a photon mass spectrum
cut-off for the renormalization of off-shell quantum electrodynamics.)

Once the divergent integrals have been regulated, one can apply the
usual techniques of multiplicative renormalization in the context of
the Feynman diagram formalism obtained in
\sect{sect:interacting:feynman}. However, further discussion of
renormalization is beyond the scope of the present paper. An
intriguing direction for future exploration is the development of a
complete regularization and renormalization program based on a
physically motivated formulation of spacetime interactions. This would
be consistent with the first-quantized approach of considering the
series expansion to be the primary representation of the physical
situation of the scattering amplitude, rather than a perturbative
approximation to a non-perturbative Lagrangian formulation.

A potentially more serious issue is whether, even after
renormalization, series such as that in \eqn{eqn:Y1} converge at all.
However, Dyson's classic argument against convergence \cite{Dyson52}
is based on the conception of traditional quantum electrodynamics,
where such series result from perturbation expansion. In the present
first-quantized formalism, Dyson's argument might simply imply that
the traditional formalism, and arguments from it, are not always
applicable.

Actually, it is not the convergence of series for probability
amplitudes, such as \eqn{eqn:Y1}, that is really important. Rather,
the real issue is whether there is a well-defined limit as $N \to
\infty$ for physically testable probabilities such as given by
\begin{equation*}
    \frac{\sqr{\bra{\alpha_{out}}\opS^{(N)}\ket{\psi_{in}}}}
         {\bra{\psi_{in}}\opS^{(N)\ddag}\opS^{(N)}\ket{\psi_{in}}} \,,
\end{equation*}
where $\opS^{(N)}$ is the result of summing \eqn{eqn:Z5} to $N$th
order, $\ket{\psi_{in}}$ is a properly normalized multiparticle in
state and $\ket{\alpha_{out}}$ is a member of a complete basis for
multiparticle out states. Quantities such as this for, say, QED
produce values that agree with experiment for large $N$. If it turns
out that they do diverge for very large $N$, this just means that
there is some mechanism in the real universe that suppresses the
interference effect of interaction graphs with very large $N$,
producing a finite cutoff of the series in \eqn{eqn:Z5}.

Indeed, from this perspective, the Lagrangian and Hamiltonian
formulations could be viewed as the approximations, obtained by
assuming the summing of series for $N \to \infty$. In the end, the
problem of divergences might even be seen as an artifact of the
conventional second-quantized Lagrangian formulation itself, rather
than of its perturbation expansion. Clearly this is an area that bears
continued exploration.

\section{Conclusion} \label{sect:conclusion}

Spacetime approaches to relativistic quantum mechanics have been
developed along a number of different threads in the literature, from
the early work on proper time formalisms by Schwinger and others
\cite{schwinger51, fock37, nambu50}, to the equally early work of
Stueckelberg \cite{stueckelberg41,stueckelberg42} and the
parametrized relativistic quantum theory it inspired \cite{collins78,
fanchi78, fanchi83, fanchi86, fanchi93, horwitz73, piron78}, to the
path integral approach introduced by Feynman \cite{feynman50,
morette51, cooke68, barut89} and its application to quantum gravity
\cite{teitelboim82, hartle86, hartle95}, to the worldline formalism
obtained as the infinite-tension limit of string theory \cite{bern88,
bern91, bern92, strassler92, schmidt93, schmidt94, schubert01} and its
relation to the typically first-quantized approach to interaction
taken in string theory \cite{green87}. The formalism presented in
the previous sections can be seen as a foundation underlying all these
approaches.

A particularly significant additional result is the derivation of
on-shell particles and antiparticle states as the infinite time limit
of free particle states. This provides a connection between off-shell
parametrized spacetime quantum theories \cite{frastai95, horwitz81,
horwitz82, shnerb93, land98b} and traditional on-shell quantum field
theory. It also suggests the intiguing possibility that, while real
particles are likely on-shell to a very high degree of approximation,
there may be testable consequences to this approximation not being
exact.

The foundation presented here provides a number of interesting 
avenues for exploration in future publications.

The approach can be readily extended to incorporate path integral
representations for non-scalar particles \cite{bordi80, henneaux82,
barut84, mannheim85, forte05}. It can also handle massless particles,
though it is not so straightforward to deal properly with the
resulting gauge symmetries \cite{shnerb93} and non-Abelian
interactions.

Further, an important payoff of the spacetime path formalism is the
intuitive grounding it gives to the theory, as opposed to the somewhat
arbitrary mathematical justifications for introducing fields in
traditional quantum field theory. Moreover, the formalism for
interacting spacetime paths provides interesting possibilities for
addressing the issues of regularization and renormalization (which is
all the more important because of the first-quantized nature of the
formalism).

Finally, a natural interpretational framework for the formalism is the
consistent histories approach to quantum theory \cite{gellmann90,
griffiths84, griffiths02, omnes88, omnes94}. Particle paths can be
treated as fine-grained histories in the sense of this approach, with
coarse-grained histories corresponding to the superposition of
fine-grained states, including cosmological histories of the universe
as a whole \cite{hartle91a, hartle91b, hartle92, hartle95}.

For example, scattering probabilities can be considered to represent
the probabilities of decohering alternative coarse-grained spacetime
histories for the scattering process. Probabilities can even be given
to decohering cosmological histories of the universe
\cite{seidewitz06b}. Such an interpretation also provides for a
natural way to see how the macroscopic classical view of the universe
emerges from the more detailed quantum description, rather than
viewing quantum physics as a ``quantization'' of a classical
description (see, for example, \refcite{halliwell03}), just as one
would wish from a foundational quantum theory.

\begin{acknowledgments}
    I would like to thank George Baker for his encouragement on an
    early version of this work and express my great appreciation to
    Philip Johnson and Lawrence Horwitz for extensive comments and
    discussion on subsequent versions.
\end{acknowledgments}

    \appendix

\section{Form of the Phase Functional} \label{app:phase}

\begin{theorem*}
    The phase functional $S$ must have the form
    \begin{equation} \label{eqn:D1A}
        S[\qdot] = \int^{\lambda_{1}}_{\lambdaz} \dl' \, 
                       L[\qdot;\lambda'] \,,
    \end{equation}
    where the parametrization domain for $\qdot$ is
    $[\lambdaz,\lambda_{1}]$ and $L[\qdot;\lambda]$ depends only on
    $\qdot$ and its higher derivatives evaluated at $\lambda$.
\end{theorem*}
\begin{proof}
In 
\begin{equation*}
    S[\qdot;\lambda',\lambdaz] = S[\qdot;\lambda',\lambda] +
                                ÊS[\qdot;\lambda,\lambdaz] \,,
\end{equation*}
consider $\lambda' = \lambda + \delta\lambda$, for infinitesimal
$\delta\lambda$:
\begin{equation*}
    \begin{split}
	S[\qdot;\lambda+\delta\lambda,\lambdaz] 
	    &= S[\qdot;\lambda+\delta\lambda,\lambda] + 
	       S[\qdot;\lambda,\lambdaz] \\
	    &\approx \delta\lambda 
               \left.
                   \pderiv{S[\qdot;\lambda',\lambda]}{\lambda'}
	       \right|_{\lambda' = \lambda} +
               S[\qdot;\lambda,\lambdaz] \,,
    \end{split}
\end{equation*}
or
\begin{equation*}
    \frac{S[\qdot;\lambda+\delta\lambda,\lambdaz] - S[\qdot;\lambda,\lambdaz]} 
         {\delta\lambda} \approx
        \left.
            \pderiv{S[\qdot;\lambda',\lambda]}{\lambda'}
        \right|_{\lambda' = \lambda} \,.
\end{equation*}
Taking the limit $\delta\lambda \to 0$ then gives
\begin{equation} \label{eqn:C0}
    \pderiv{S[\qdot;\lambda,\lambdaz]}{\lambda} = L[\qdot;\lambda] \,,
\end{equation}
where
\begin{equation*}
    L[\qdot;\lambda] \equiv 
        \left.
            \pderiv{S[\qdot;\lambda',\lambda]}{\lambda'}
        \right|_{\lambda' = \lambda} \,.
\end{equation*}
Now, the functional $L$ depends only on $\qdot$ and $\lambda$, not
$\lambdaz$. Therefore, integrate \eqn{eqn:C0} over $\lambda$, with the
initial condition $S[\qdot;\lambdaz,\lambdaz] = 0$, to get
\begin{equation*}
    S[\qdot;\lambda,\lambdaz] = \int^{\lambda}_{\lambdaz} \dl' \, 
                                L[\qdot;\lambda'] \,,
\end{equation*}
which is just \eqn{eqn:D1A}.

Further, by definition $S[\qdot;\lambda,\lambdaz]$ only depends on
values of $\qdotmu$ between $\lambdaz$ and $\lambda$. Therefore,
$S[\qdot;\lambda+\delta\lambda,\lambda] \approx L[\qdot;\lambda] 
\delta\lambda$ should only depend on $\qdot$ infinitesimally close to 
$\lambda$. As $\delta\lambda \to 0$, this effectively limits 
$L[\qdot;\lambda]$ to depend only on $\qdot$ and its derivatives 
evaluated at $\lambda$.

\end{proof}

\section{Evaluation of the Path Integral} \label{app:path}

\begin{theorem*}
    The path integral
    \begin{equation} \label{eqn:F1A}
        \kernel =
            \eta[\lambda]
            \intDfour q\, 
                \delta^{4}(q(\lambda(1)) - x)
                \delta^{4}(q(\lambda(0)) - \xz)
                    \exp \left(
                        \mi\int^{\lambda}_{\lambdaz}\dl' \,
                            \left[
                                \frac{1}{4}\qdotsq(\lambda') - m^{2}
                            \right]
                    \right) \,.
    \end{equation}
    may be evaluated to get
    \begin{equation} \label{eqn:HA}
        \kernel = \kerneld 
            \equiv (2\pi)^{-4} \intfour p \, \me^{\mi p\cdot(x - \xz)}
                   \me^{-\mi(\lambda - \lambdaz)(p^{2} + m^{2})} \,.
    \end{equation}
\end{theorem*}
\begin{proof}
The path integral in \eqn{eqn:F1A} may be defined as
\begin{equation*}
    \kernel = \lim_{N \to \infty} \propbar \,,
\end{equation*}
where
\begin{multline} \label{eqn:F3}
    \propbar \equiv
          \bar{\eta}(\bar{\lambda}_{0},\ldots,\bar{\lambda}_{N})
          \int \dfour \bar{q}_{0} \cdots \dfour \bar{q}_{N} \, 
          \delta^{4}(\bar{q}_{N} - x)
          \delta^{4}(\bar{q}_{0} - \xz) \\
          \times
          \exp\left(
              \mi\sum_{j=1}^{N}\Delta\bar{\lambda}_{j}
                  (\frac{1}{4}\bar{\qdot}_{j}^{2} - m^{2})
          \right) \,,
\end{multline}
$\bar{\eta}(\bar{\lambda}_{0},\ldots,\bar{\lambda}_{N}) \to
\eta[\lambda]$ as $N \to \infty$ and the $N$-point discrete
approximations to the functions $\lambda(s)$ and $q(\lambda(s))$ are
given by
\begin{equation*}
    \bar{\lambda}_{j} = \lambda(j/N)
\end{equation*}
and
\begin{equation} \label{eqn:F3a}
    \bar{q}_{j} = q(\bar{\lambda}_{j}) \,,
\end{equation}
for $j = 0, \ldots , N$. The $\lambda$ integral is approximated by a
summation with
\begin{equation*}
    \Delta\bar{\lambda}_{j} 
        \equiv \bar{\lambda}_{j} - \bar{\lambda}_{j-1}
\end{equation*}
and
\begin{equation} \label{eqn:F4}
    \bar{\qdot}_{j} 
        \equiv (\bar{q}_{j} - \bar{q}_{j-1})/\Delta\bar{\lambda}_{j} \,,
\end{equation}
for $j = 1, \ldots , N$. 

To compute the path integral, insert the product of Gaussian integrals
\begin{equation*}
    \prod_{j=1}^{N} \mi 
        \left(
            \frac{\Delta\bar{\lambda}_{j}}{\pi}
        \right)^{2}
        \intfour \bar{p}_{j} \, 
            \me^{-\mi\Delta\bar{\lambda}_{j}\bar{p}_{j}^{2}} 
    = 1
\end{equation*}
into the N-point approximation of \eqn{eqn:F3} to get
\begin{multline*}
    \propbar =
         \bar{\xi}(\bar{\lambda}_{0},\ldots,\bar{\lambda}_{N})
	 \int \dfour \bar{q}_{0} \cdots \dfour \bar{q}_{N}
         \int \dfour \bar{p}_{1} \cdots \dfour \bar{p}_{N} \,
         \delta^{4}(\bar{q}_{N} - x) 
         \delta^{4}(\bar{q}_{0} - \xz) \\
         \times
         \exp\left(
             \mi\sum_{j=1}^{N}\Delta\bar{\lambda}_{j}
             (-\bar{p}_{j}^{2} + 
                   \frac{1}{4}\bar{\qdot}_{j}^{2} - m^{2})
         \right) \,,
\end{multline*}
where
\begin{equation*}
    \bar{\xi}(\bar{\lambda}_{0},\ldots,\bar{\lambda}_{N}) \equiv 
        \left[ \prod_{j=1}^{N} \mi 
            \left(
                \frac{\Delta\bar{\lambda}_{j}}{\pi}
            \right)^{2}
        \right]
        \bar{\eta}(\bar{\lambda}_{0},\ldots,\bar{\lambda}_{N}) \,.
\end{equation*}
Inside the $\bar{p}_{j}$ integrals, make the change of variables
$\bar{p}_{j} \to \bar{p}_{j} - \frac{1}{2}\bar{\qdot}_{j}$, so that
\begin{equation} \label{eqn:F6}
        \sum_{j=1}^{N}\Delta\bar{\lambda}_{j}(-\bar{p}_{j}^{2} + 
        \frac{1}{4}\bar{\qdot}_{j}^{2} - m^{2})
	    \to \sum_{j=1}^{N}\Delta\bar{\lambda}_{j}(-\bar{p}_{j}^{2} + 
	         \bar{p}_{j}\cdot\bar{\qdot}_{j} - 
		 \frac{1}{4}\bar{\qdot}_{j}^{2} +
		 \frac{1}{4}\bar{\qdot}_{j}^{2} -m^{2}) \\
	    = \sum_{j=1}^{N}\Delta\bar{\lambda}_{j}
                  [\bar{p}_{j}\cdot\bar{\qdot}_{j} - 
                      (\bar{p}_{j}^{2} + m^{2})] \,.
\end{equation}
Now, using \eqn{eqn:F4},
\begin{equation*}
    \sum_{j=1}^{N}\Delta\bar{\lambda}_{j}\bar{p}_{j}\bar{\qdot}_{j}
        = \sum_{j=1}^{N}\bar{p}_{j}\cdot(\bar{q}_{j}-\bar{q}_{j-1})
        = \bar{p}_{N}\cdot\bar{q}_{N} - \bar{p}_{1}\cdot\bar{q}_{0} -
            \sum_{j=1}^{N-1}(\bar{p}_{j+1}-\bar{p}_{j})\cdot\bar{q}_{j}
\end{equation*}
(this is essentially just integration by parts within the
approximation to the path integral). But, for each $\bar{q}_{j}$, $j =
1, \ldots, N-1$,
\begin{equation*}
    \intfour \bar{q}_{j} \, 
    \me^{-\mi(\bar{p}_{j+1}-\bar{p}_{j})\cdot\bar{q}_{j}}
        = (2\pi)^{4} \delta^{4}(\bar{p}_{j+1}-\bar{p}_{j}) \,,
\end{equation*}
so, integrating over the $\bar{p}_{j}$ for $j = 2, 3, \ldots, N$ 
gives $\bar{p}_{j+1} = \bar{p}_{j}$. Therefore,
\begin{equation} \label{eqn:G}
    \begin{split}
	\propbar 
            &= \begin{aligned}[t]
                   &\bar{\xi}(\bar{\lambda}_{0},\ldots,
                   \bar{\lambda}_{N})
                   \int \dfour\bar{q}_{0}\,
                   \dfour\bar{q}_{N} \,
                   \delta^{4}(\bar{q}_{N} - x)
                   \delta^{4}(\bar{q}_{0} - \xz) \\
		   &\qquad \times
                   \intfour \bar{p}_{1} \, (2\pi)^{4(N-1)}
                   \me^{\mi\bar{p}_{1}\cdot(\bar{q}_{N} - \bar{q}_{0})}
                   \exp\left(-\mi\sum_{j=1}^{N}\Delta\bar{\lambda}_{j}
                            (\bar{p}_{1}^{2} + m^{2})\right) 
               \end{aligned} \\
            &= (2\pi)^{-4}
               \bar{\zeta}(\bar{\lambda}_{0},\ldots,\bar{\lambda}_{N})
               \intfour p \,
               \me^{\mi p\cdot(x - \xz)} 
               \me^{-\mi(\bar{\lambda}_{N} - \bar{\lambda}_{0})
                   (p^{2} + m^{2})} \,,
    \end{split}
\end{equation}
where
\begin{equation*}
    \bar{\zeta}(\bar{\lambda}_{0},\ldots,\bar{\lambda}_{N})
        \equiv (2\pi)^{4N} 
               \bar{\xi}(\bar{\lambda}_{0},\ldots,\bar{\lambda}_{N})
             = \left[
                   \prod_{j=1}^{N}\mi(4\pi\Delta\bar{\lambda}_{j})^{2}
               \right]
               \bar{\eta}(\bar{\lambda}_{0},\ldots,\bar{\lambda}_{N}) \,.
\end{equation*}

Now set the normalization factor
\begin{equation*}
    \bar{\eta}(\bar{\lambda}_{0},\ldots,\bar{\lambda}_{N})
        = \prod_{j=1}^{N} (-\mi)(4\pi\Delta\bar{\lambda}_{j})^{-2} \,.
\end{equation*}
Then $\bar{\zeta}(\bar{\lambda}_{0},\ldots,\bar{\lambda}_{N}) = 1$, so
we can take the limit $N \to \infty$ of \eqn{eqn:G} to get
\eqn{eqn:HA}.
\end{proof}

    \bibliography{../../RQMbib}
    
\end{document}